\documentclass[apj]{emulateapj}
\usepackage{lscape}

\newcommand{\oi}{[\textrm{O}~\textsc{i}]}
\newcommand{\oii}{[\textrm{O}~\textsc{ii}]}
\newcommand{\oiii}{[\textrm{O}~\textsc{iii}]}
\newcommand{\oiv}{[\textrm{O}~\textsc{iv}]}
\newcommand{\neii}{[\textrm{Ne}~\textsc{ii}]}
\newcommand{\neiii}{[\textrm{Ne}~\textsc{iii}]}

\shorttitle{Black Hole Growth and Star Formation in Seyfert Galaxies}
\shortauthors{Diamond-Stanic et al.}

\begin{document}
\slugcomment{Accepted for publication in ApJ}

\title{The Relationship Between Black Hole Growth \\ and Star Formation
  in Seyfert Galaxies}

\author{Aleksandar M. Diamond-Stanic\altaffilmark{1,2}, George
  H. Rieke\altaffilmark{3}} 

\altaffiltext{1}{Center for Astrophysics and Space Sciences,
  University of California, San Diego, La Jolla, CA, 92093}
\altaffiltext{2}{Center for Galaxy Evolution Fellow}
\altaffiltext{3}{Steward Observatory, University of Arizona, Tucson,
  AZ, 85721}

\begin{abstract}
We present estimates of black hole accretion rates and nuclear,
extended, and total star-formation rates for a complete sample of
Seyfert galaxies.  Using data from the Spitzer Space Telescope, we
measure the active galactic nucleus (AGN) luminosity using the
\oiv~$\lambda25.89~\mu$m emission line and the star-forming luminosity
using the 11.3~$\mu$m aromatic feature and extended 24~$\mu$m
continuum emission.  We find that black hole growth is strongly
correlated with nuclear ($r<1$~kpc) star formation, but only weakly
correlated with extended ($r>1$~kpc) star formation in the host
galaxy.  In particular, the nuclear star-formation rate (SFR) traced
by the 11.3~$\mu$m aromatic feature follows a relationship with the
black hole accretion rate (BHAR) of the form $SFR\propto
\dot{M}_{BH}^{0.8}$, with an observed scatter of 0.5~dex.  This
SFR--BHAR relationship persists when additional star formation in
physically matched $r=1$~kpc apertures is included, taking the form
$SFR\propto \dot{M}_{BH}^{0.6}$.  However, the relationship becomes
almost indiscernible when total SFRs are considered.  This suggests a
physical connection between the gas on sub-kpc and sub-pc scales in
local Seyfert galaxies that is not related to external processes in
the host galaxy.  It also suggests that the observed scaling between
star formation and black hole growth for samples of AGNs will depend
on whether the star formation is dominated by a nuclear or extended
component.  We estimate the integrated black hole and bulge growth
that occurs in these galaxies and find that an AGN duty cycle of
5--10\% would maintain the ratio between black hole and bulge masses
seen in the local universe.
\end{abstract}


\section{Introduction}\label{sec:intro}

The discovery of correlations between the masses of supermassive black
holes and the properties of galaxy bulges such as mass \citep{kor95,
  mag98, mar03, har04} and velocity dispersion \citep{fer00, geb00,
  tre02} suggests a connection between the processes that regulate the
growth of the central black hole and the galaxy bulge.  Some models
have explained this connection via energetic feedback from the
accreting black hole \citep{sil98, wyi03, dim05, som08}, which is
often assumed to be triggered by galaxy mergers
\citep[e.g.,][]{kau00,hop05}.  While such violent, merger-driven
activity may be important for fueling luminous quasars, secular
processes in the host galaxy \citep[e.g.,][]{hop06,jog06} may be
sufficient to explain the black hole accretion rates (BHARs) of
lower-luminosity active galactic nuclei (AGNs) such as Seyfert
galaxies.

Important insights regarding the black hole--galaxy bulge connection
and mechanisms for fueling AGNs can be obtained by studying the
stellar populations of AGN host galaxies.  There is evidence that
circumnuclear stellar populations around AGNs include a relatively
young component \citep[e.g.,][]{cid01,dav07,rif09} and that higher
luminosity AGNs tend to be associated with younger stellar populations
\citep[e.g.,][]{kau03,wil07}.  This connection may be driven by the
fact that both black hole growth and star formation occur when large
amounts of gas are injected into the central region of a galaxy
\citep[e.g.,][]{san88,bar91,sto01}, although a mechanism is necessary
to transport gas from $r\sim100$~pc down to sub-pc scales, regardless
of the external fuel supply \citep[e.g.,][]{shl90,wad04,esc07,hop10}.
A more direct connection between nuclear star formation and AGN
activity has also been proposed where mass loss from evolved stars
\citep[e.g.,][]{nor88,cio07} or angular-momentum loss associated with
supernova-generated turbulence \citep[e.g.,][]{von93,hob11} supplies
the necessary material to the black hole accretion disk.

Measurements of ongoing star formation in AGN host galaxies are useful
for constraining the above models, but standard star-formation rate
(SFR) diagnostics such as the ultraviolet (UV) continuum, the
H$\alpha$ emission line, or the infrared (IR) continuum
\citep[e.g.,][]{ken98} are often contaminated by the AGN itself.  The
mid-IR aromatic (also known as PAH) features offer a solution because
they probe the strength of the UV radiation field in
photo-dissociation regions near young, massive stars
\citep[e.g.,][]{pee04,tie05,smi07a}.  While high-energy photons or
shocks associated with the AGN could destroy the molecular carriers of
the aromatic features \citep[e.g.,][]{voi92,gen98}, \citet{dia10}
showed that the 11.3~$\mu$m feature remains a valid measure of the SFR
for the nuclear environments of local Seyfert galaxies.

Previous studies that have used the aromatic features to assess the
level of star formation in AGN host galaxies have found it to be
correlated with the AGN luminosity
\citep[e.g.,][]{ima04,sch06,shi07,net07,lut08,shi09}.  This is broadly
consistent with models where star formation and AGN activity are both
triggered by an external supply of cold gas or models where nuclear
star formation fuels subsequent AGN activity.  However, it is unclear
how this empirical relationship behaves for a more complete sample of
AGNs, how it varies as function of scale in the galaxy (e.g., nuclear
v. extended star formation), or to what extent there are differences
as a function of AGN luminosity or obscuration \citep[e.g.,][]{lut10}.

In this paper, we extend the study of SFRs in AGN host galaxies to a
complete sample of Seyfert galaxies drawn from the revised
Shapley--Ames galaxy catalog \citep[RSA,][]{san87,mai95,ho97}.  All
galaxies have nuclear mid-IR spectra from the Infrared Spectrograph
\citep[IRS,][]{hou04} and imaging from the Multiband Imaging
Photometer for Spitzer \citep[MIPS,][]{rie04} onboard the Spitzer
Space Telescope \citep{wer04}, and we are able to treat nuclear
($<1$~kpc) and extended regions separately.  This spatially resolved
information allows us to assess the extent to which star formation and
black hole accretion activity are physically connected.  We present
measurements of BHARs based on the \oiv~$\lambda25.89~\mu$m emission
line \citep[e.g.,][]{dia09,rig09}, nuclear SFRs based on the
11.3~$\mu$m aromatic feature \citep[e.g.,][]{dia10}, and extended SFRs
based on 24~$\mu$m flux \citep[e.g.,][]{rie09}.  We explore the
relationships among these quantities and the constraints they place on
models of black hole growth and galaxy evolution.

\section{Sample, Data, and Measurements}\label{sec:data}

We consider the RSA Seyfert sample analyzed by \citet{dia09}, which
includes the 89 Seyferts brighter than $B_T=13$ from \citet{mai95} and
\citet{ho97}.  This galaxy-magnitude-limited sample is unique in its
sensitivity to low-luminosity and highly obscured AGNs \citep{mai95,
  ho97}.  The median distance of the sample is 22~Mpc, where the
3.7\arcsec~slit width of the IRS Short-Low (SL) module subtends
390~pc.  Distances and Seyfert types are compiled by \citet{dia09}.

For the analysis of BHARs, we use \oiv~$\lambda25.89~\mu$m
emission-line fluxes from \citet{per10}, which are based on data from
the Long-High IRS module.  For the nine galaxies without
\oiv\ measurements from \citet{per10}, we use \oiv\ upper limits
\citet{dia09}, which are based on data from the Long-Low IRS module.
When considering $\oiv/\neii$ ratios, we use \neii\ fluxes from
\citet{per10}, which are based on data from the Short-High IRS module.
Among the 84 sources with 11.3~$\mu$m aromatic feature measurements
(see below), we focus our analysis on the 74 galaxies with
$\oiv/\neii>0.15$, including 12 sources with \oiv\ upper limits that
are consistent with this ratio.  For these galaxies the contribution
of star formation to \oiv\ is sub-dominant \citep[$<30\%$,][]{per10},
so we are able to use their \oiv\ fluxes to robustly estimate BHARs.

For the analysis of nuclear SFRs (Section~\ref{sec:nucsfr}), we use
measurements of the 11.3~$\mu$m aromatic feature and the
\neii~$\lambda12.81~\mu$m emission line based on data from the IRS SL
module.  We used the IRS SL module (rather than the SH module) for
\neii\ SFRs to enable comparisons with aromatic-based SFRs in the same
aperture.  Following \citet{dia10}, one-dimensional spectra were
extracted using CUBISM \citep{smi07a} with small apertures
($3.6\arcsec\times7.2\arcsec$) designed to isolate nuclear emission,
and spectral fitting was performed with PAHFIT \citep{smi07b}.  For
three sources (NGC3031, NGC3783, NGC4151), we did not obtain an
adequate continuum fit near 11.3~$\mu$m, so we defined the local
continuum using a power-law fit in the $\lambda=10.75$--10.9~$\mu$m
and $\lambda=11.65$--11.85~$\mu$m regions.  Our measurements are
compiled in Table~\ref{tab:data}.  We exclude two sources (CIRCINUS,
NGC1068) with saturated SL2 data and three sources (NGC777, NGC4168,
NGC4472) without 11.3~$\mu$m aromatic feature or
\oiv~$\lambda25.89~\mu$m detections, leaving a sample of 84 galaxies.

Aperture corrections are necessary to determine total nuclear fluxes
because the above measurements only cover the central portion
($1.2\times2.3~\lambda/D$) of the point-spread function function (PSF)
at $\lambda\approx12~\mu$m.  Using the IRAC channel 4 (8.0~$\mu$m)
PSF, scaled to match the expected beam size at $\lambda=11.3~\mu$m
($\lambda=12.8~\mu$m), we find that our extraction aperture contains
only 51.9\% (45.3\%) of the total flux from a point source.  When
determining SFRs, we therefore apply an aperture correction of 1.93
(2.21) to the 11.3~$\mu$m aromatic feature (\neii) fluxes.

Using MIPS 24~$\mu$m images, we also consider the star formation in
regions of each galaxy that are outside the IRS SL
$3.6\arcsec\times7.2\arcsec$ extraction aperture.  We measure
24~$\mu$m fluxes in circular apertures with radii $r=2.9\arcsec$,
$r=1$~kpc, and $r=r_{\textnormal{\scriptsize{galaxy}}}$.  For each
galaxy we attribute all the 24~$\mu$m flux measured in the
$2.9\arcsec$ aperture (diameter
$\approx\lambda/D$)\footnote{$r=(3.6\arcsec\times7.2\arcsec/\pi)^{0.5}=2.9\arcsec$}
to a central point source (i.e., unresolved emission from both the AGN
and nuclear star formation) and we use aperture corrections based on
the MIPS 24~$\mu$m PSF to determine the point-source contribution in
the larger apertures.  We exclude these contributions and convert the
remaining fluxes into SFRs using the \citet{rie09} calibration.  The
choice of $r=1$~kpc matches the area subtended by the
$3.6\arcsec\times7.2\arcsec$ aperture for the most distant galaxies
and corresponds to the smallest physical aperture that can be used for
the whole sample\footnote{$r=2.9\arcsec$ corresponds to $r=0.97$~kpc
  at $D=70$ Mpc.  The only galaxy with $D>70$~Mpc is Mrk 509
  \citep{dia09}.}.

\begin{figure}[!t]
\begin{center}
\hspace{-1cm}
\includegraphics[angle=0,scale=.47]{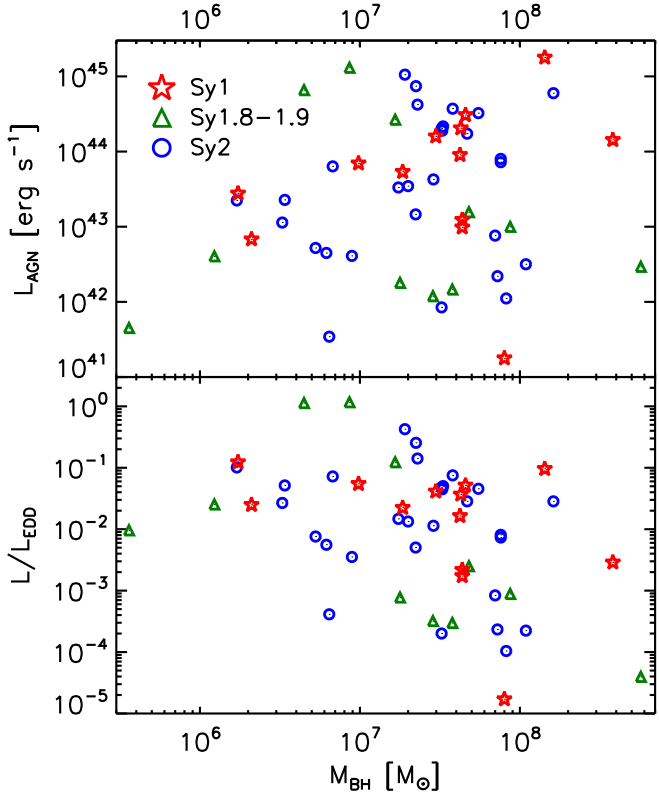}
\caption{The demographics of AGN luminosity, Eddington ratio, and
  black hole mass for the RSA Seyfert sample.  The sample includes
  low- to moderate-luminosity AGNs with masses generally below
  $10^{8}~M_{\odot}$.  The most rapidly accreting sources with
  $L/L_{Edd}>0.1$ all have $M_{BH}<3\times10^{7}~M_{\odot}$.}
\label{fig:ml}
\end{center}
\end{figure}

\section{Black Hole Accretion Rates}

The \oiv\ line has been shown to be an accurate tracer of AGN
intrinsic luminosity \citep{mel08a,rig09,dia09}.  The line is
dominated by the AGN unless the IR luminosity associated with SF
exceeds the AGN intrinsic luminosity by an order of magnitude
\citep{per10}.  We use the calibration of \citet{rig09} to convert
between \oiv\ luminosity and AGN intrinsic luminosity,
$L_{\textnormal{\scriptsize{AGN}}}=L_{\scriptsize{\oiv}}\times2550$,
which has an rms scatter of 0.4 dex.  Assuming a radiative efficiency
$\eta=0.1$ ($L_{\textnormal{\scriptsize{AGN}}}=\eta\dot{M}_{BH}c^2$),
this is equivalent to the following:
\begin{equation}
\dot{M}_{BH} (M_{\odot}~\textnormal{yr}^{-1}) = 1.7\times10^{-9} L
(\oiv, L_{\odot})
\label{eq:oiv}
\end{equation}
We note that there are theoretical expectations that the radiative
efficiency may drop at both high \citep[$L/L_{Edd}>1$, e.g.,][]{abr88}
and low \citep[$L/L_{Edd}<0.01$, e.g.,][]{nar95} accretion rates due
to advection of matter onto the black hole.  The latter regime is
relevant for our sample, such that equation~\ref{eq:oiv} may
underestimate the true mass accretion rate for sources with small
$L/L_{Edd}$ values.

To assess the demographics of the sample in terms of the ratio of the
AGN intrinsic luminosity to the Eddington luminosity,
L$_{\textnormal{\scriptsize{EDD}}}=1.3\times10^{46}(M_{BH}/10^{8}M_{\odot})$~erg~s$^{-1}$,
we gathered estimates of black hole mass from the literature based on
high-resolution gas, stellar, or maser dynamics (5 objects),
reverberation mapping (10 objects), and bulge velocity dispersion (46
objects).  The values of AGN intrinsic luminosity, black hole mass,
and Eddington ratio for the sample are shown in Figure~\ref{fig:ml}
and compiled in Table~\ref{tab:data}.  Most objects fall in the range
$L_{AGN}=10^{42}$--$10^{45}$~erg~s$^{-1}$,
$M_{BH}=10^6$--$10^{8}~M_{\odot}$, and $L/L_{Edd}=10^{-4}$--1.

\section{Nuclear Star-Formation Rates}\label{sec:nucsfr}

The mid-IR aromatic features \citep[e.g.,][]{pee04,smi07a,cal07}, the
\neii\ line \citep[e.g.,][]{ho07,diaz10}, and the 24~$\mu$m continuum
luminosity \citep[e.g.,][]{cal07,rie09} can all be used as tracers of
the SFR for normal star-forming galaxies.  However, when a galaxy
contains a central AGN, dust heated by the AGN will likely dominate
the 24~$\mu$m continuum \citep[e.g.,][]{bra06}, ionizing photons from
the AGN can contribute significantly to
\neii\ \citep[e.g.,][]{gro06,per10}, and high-energy photons or shocks
associated with the AGN may destroy or modify the molecules that
produce the mid-IR aromatic features
\citep[e.g.,][]{voi92,odo09,dia10}.  Nonetheless, \citet{dia10} showed
that the 11.3~$\mu$m aromatic feature is robust to the effects of AGN-
and shock-processing, and \citet{mel08b} outlined a method to
determine the star-formation contribution to \neii\ for AGNs.  In this
section, we evaluate the merit of the 11.3~$\mu$m aromatic feature and
the \neii\ emission line for estimating nuclear SFRs of AGN host
galaxies.

\subsection{SFR Calibrations}

We used the \citet{rie09} star-forming galaxy templates to convert the
11.3~$\mu$m aromatic feature and \neii\ emission-line strengths into
SFRs.  Based on spectral decompositions with PAHFIT, we determined the
strength of these features for the templates in the
$L_{IR}=10^{9.75}$--$10^{10.75}~L_{\odot}$ range, which are
appropriate for the nuclear SFRs in the sample
($<10$~M$_{\odot}$~yr$^{-1}$).  For these templates, the 11.3~$\mu$m
aromatic feature contributes $1.2\%\pm0.1\%$ of the IR luminosity,
while \neii\ contributes $0.13\%\pm0.01\%$ (above
$L_{IR}=10^{11}~L_{\odot}$ these fractions drop to $\sim0.5\%$ and
$\sim0.07\%$, respectively).  Using the \citet{rie09} calibration
between $L_{IR}$ and SFR\footnote{The \citet{rie09} calibration yields
  SFRs that are 0.66 times those from the \citet{ken98} calibration
  due to different assumptions about the stellar initial mass
  function.}, we find for this luminosity range:
\begin{equation}
SFR (M_{\odot}~\textnormal{yr}^{-1}) = 9.6\times10^{-9} L (11.3~\mu m, L_{\odot})
\label{eq:pah}
\end{equation}
\begin{equation}
SFR (M_{\odot}~\textnormal{yr}^{-1}) = 8.9\times10^{-8} L (\neii, L_{\odot})
\label{eq:ne2}
\end{equation}

We note that equation 2 of \citet{ho07} implies that
\neii\ contributes, on average, only $\sim$0.05\% of the IR luminosity
(albeit with a large scatter of 0.51~dex).  Such a small
L$_{\scriptsize{\neii}}$/L$_{IR}$ ratio is more consistent with the
\citet{rie09} templates above $L_{IR}=10^{11}~L_{\odot}$.  For lower
luminosities, \citet{tre10} note that the sample used by \citet{ho07}
includes a number of low-metallicity galaxies where \neiii\ is the
dominant Ne species \citep[e.g.,][]{wu06}, resulting in smaller
L$_{\scriptsize{\neii}}$/L$_{IR}$ ratios.  Thus for the sources
considered in this paper, our equation~\ref{eq:ne2} is more
appropriate for converting \neii\ luminosities into SFRs.

\begin{figure*}[!t]
\begin{center}
\includegraphics[angle=90,scale=.65]{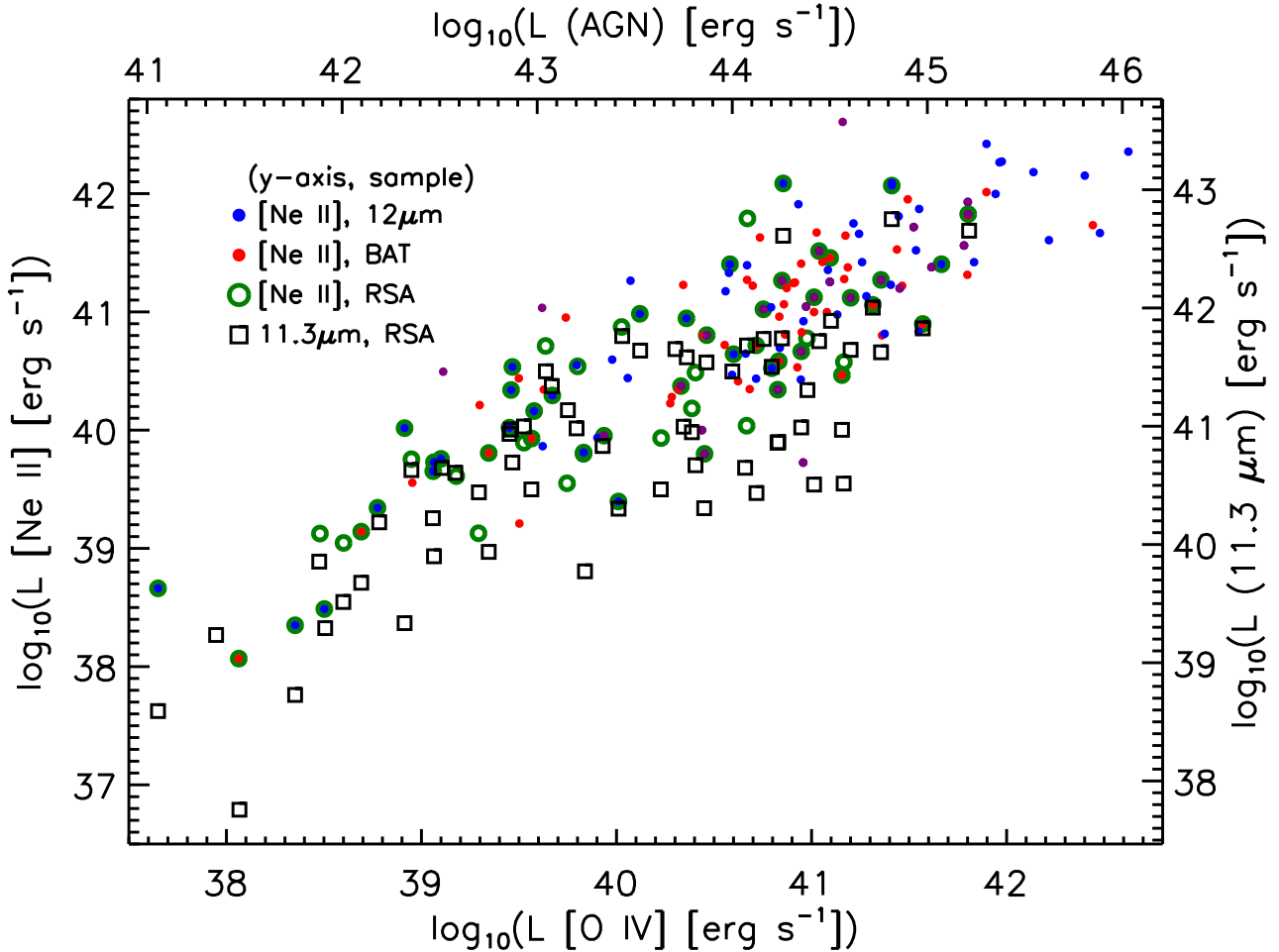}
\caption{The relationship between \neii, 11.3~$\mu$m aromatic feature,
  and \oiv\ luminosity.  We show the \neii--\oiv\ relationship for the
  12 micron, Swift-BAT, and RSA samples, as well as the
  aromatic--\oiv\ relationship for the RSA sample.  The \neii\ and
  aromatic-feature axes are normalized based on their typical ratio in
  star-forming galaxies, and the handful of starburst-dominated
  Seyferts with $\oiv/\neii<0.15$ are not plotted.  This figure
  illustrates that \neii--\oiv\ relationships is tighter than the
  aromatic--\oiv\ relationship, with the discrepancy being driven by
  sources with larger \neii-to-aromatic-feature ratios.}
\label{fig:lne2_lpah}
\end{center}
\end{figure*}

\begin{figure}[!t]
\begin{center}
\hspace{-1cm}
\includegraphics[angle=0,scale=.47]{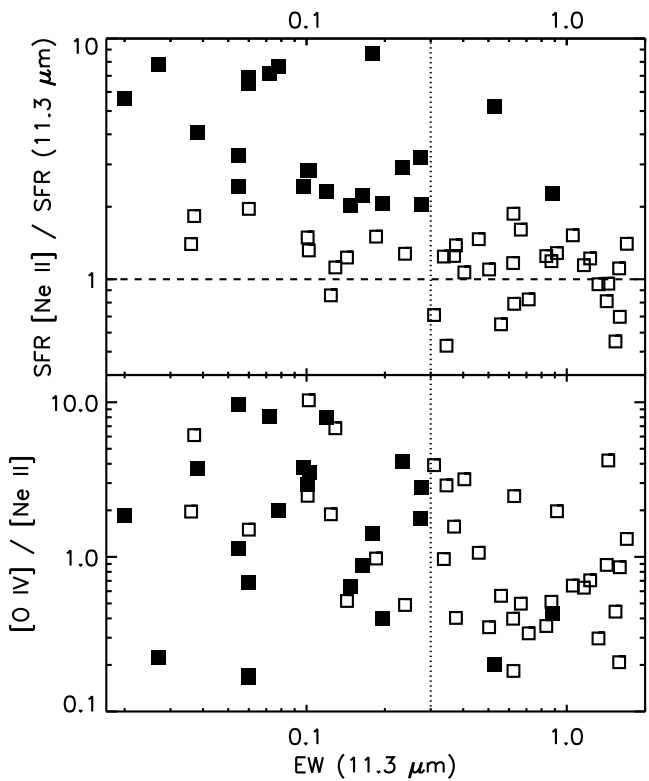}
\caption{Top panel: The ratio of SFRs derived from the \neii\ line to
  SFRs derived from the 11.3~$\mu$m aromatic feature as a function of
  EW(11.3~$\mu$m).  Sources whose \neii\ and aromatic SFRs differ by
  more than a factor of two are noted by filled symbols.  This ratio
  scatters around unity (dashed line) for sources with
  EW(11.3~$\mu$m$)>0.3$~$\mu$m (dotted line).  However, sources with
  smaller EWs tend to have systematically larger
  \neii-to-aromatic-feature ratios.  Bottom panel: The
  \oiv/\neii\ ratio as a function of EW(11.3~$\mu$m).  Sources with
  discrepant \neii\ SFRs are most commonly found in the AGN-dominated
  region (top-left) of this plot.  The two sources with
  $\oiv/\neii<0.3$ and EW(11.3~$\mu$m)$<0.1$~$\mu$m (NGC3031, NGC1275)
  are discussed in the text.}
\label{fig:o4ne2pahsfr}
\end{center}
\end{figure}

\begin{figure}[!t]
\begin{center}
\includegraphics[angle=90,scale=.38]{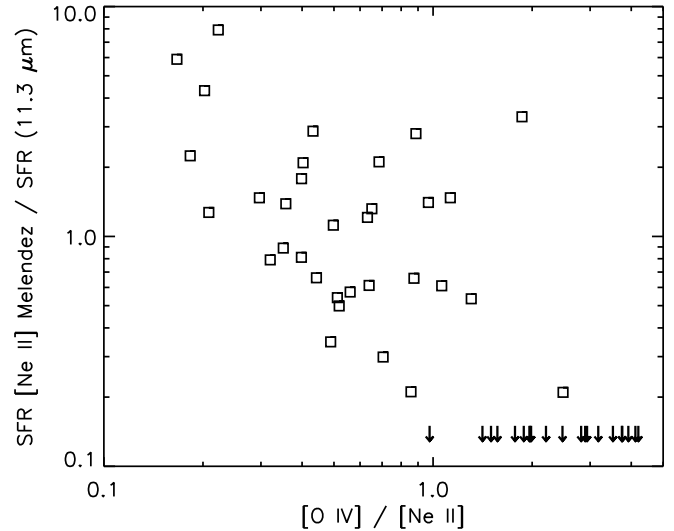}
\caption{The ratio of SFRs derived from the \neii\ line, using the
  method of \citet{mel08b} to subtract the AGN contribution to \neii,
  to SFRs derived from the 11.3~$\mu$m aromatic feature as a function
  of the \oiv/\neii\ ratio.  We find that the \citet{mel08b} method
  can overestimate SFRs for sources with $\oiv/\neii<0.3$ and
  underestimate SFRs for sources with $\oiv/\neii>1$.  }
\label{fig:melendez}
\end{center}
\end{figure}

\begin{figure*}[!t]
\begin{center}
\includegraphics[angle=90,scale=.65]{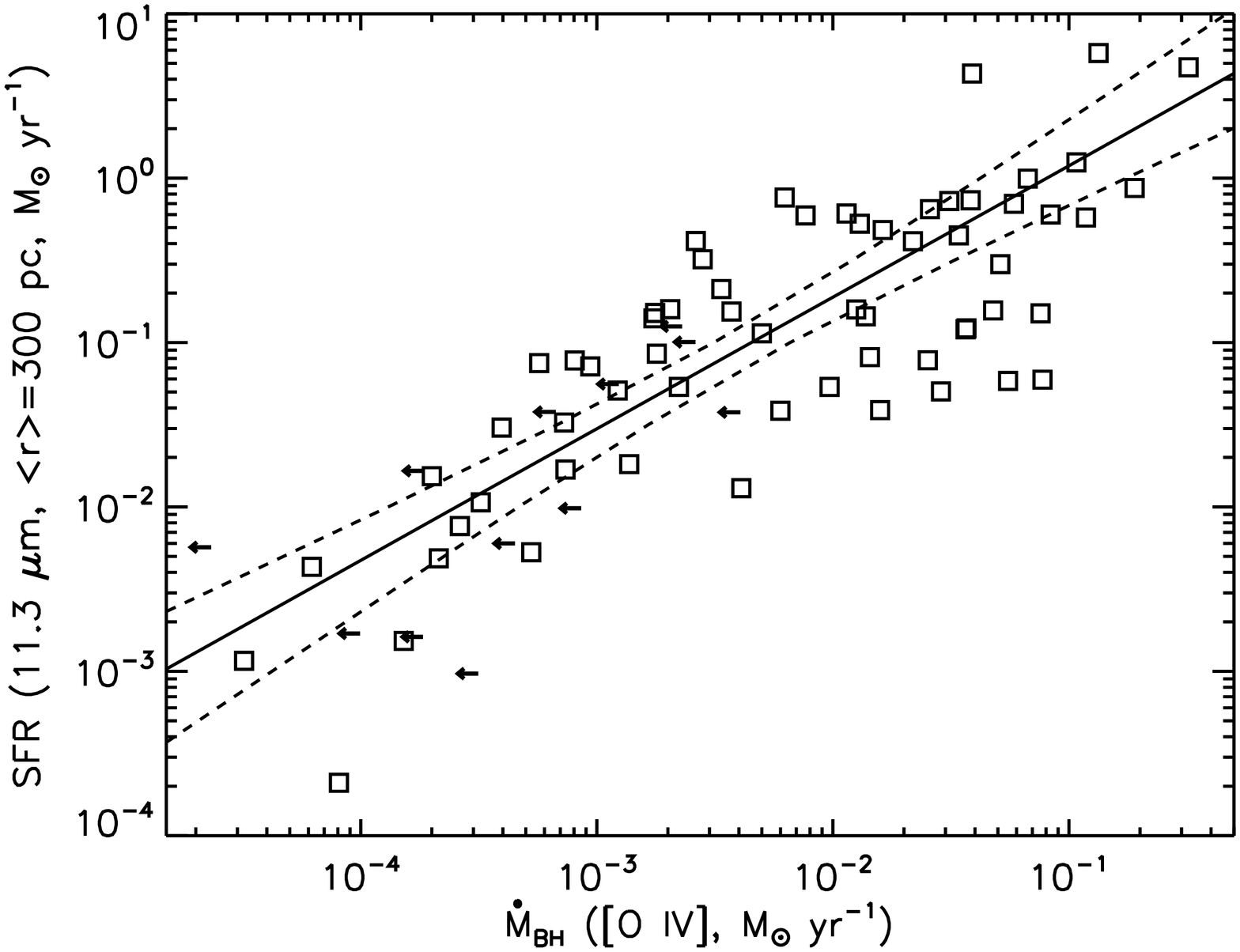}
\caption{The relationship between nuclear SFR as traced by the
  11.3~$\mu$m aromatic feature and BHAR as traced by \oiv.  Seyferts
  with high accretion rates also tend to have enhanced nuclear SFRs.
  The solid line is the best-fit relationship
  (equation~\ref{eq:sfrpah}), and the dotted lines show the 95\%
  confidence interval on the regression line.}
\label{fig:mdotsfr}
\end{center}
\end{figure*}

\subsection{Comparing Aromatic and [Ne II] SFRs}

We show the relationship between \neii, \oiv, and 11.3~$\mu$m aromatic
feature luminosities for RSA Seyferts in Figure~\ref{fig:lne2_lpah}.
For reference, we also show the \neii--\oiv\ relationship for Seyferts
from the 12~$\mu$m sample \citep[e.g.,][]{tom10} and the Swift-BAT
sample \citep[e.g.,][]{wea10} that are not starburst dominated (i.e.,
excluding the $\approx10$\% of sources with $\oiv/\neii<0.15$, see
Section~\ref{sec:data}).  This figure illustrates that there is less
scatter in the \neii--\oiv\ relationship (0.41~dex) than in the
aromatic--\oiv\ relationship (0.52~dex), implying a tighter connection
between the two quantities.  This figure also illustrates that if one
attributes all of the \neii\ flux to star formation, the \neii-based
SFR often exceeds the aromatic-based SFR.  



To investigate this behavior in more detail, we compare aromatic and
\neii\ SFRs as a function of the equivalent width (EW) of the
11.3~$\mu$m aromatic feature in the top panel of
Figure~\ref{fig:o4ne2pahsfr}.  We find good agreement between aromatic
and \neii\ SFRs for sources with EW(11.3~$\mu$m$)>0.3$~$\mu$m, with
0.2~dex of scatter around the median ratio
SFR$_{\scriptsize{\neii}}$/SFR$_{11.3~\mu m}=1.14$, consistent with
previous results from \citet{dia10} for a subset of this sample.  The
largest outlier in this regime, NGC2639, shows evidence for suppressed
aromatic features \citep{dia10}.  However, for sources with
EW(11.3~$\mu$m$)<0.3$~$\mu$m, the \neii\ SFR estimates are
systematically larger than the aromatic SFRs.  Given that the aromatic
feature EW is a proxy for the ratio of star-forming to AGN luminosity
\citep[e.g.,][]{gen98}, this behavior illustrates that the SFR
discrepancy is associated with AGN-dominated sources.

In the bottom panel of Figure~\ref{fig:o4ne2pahsfr}, we show that the
sources with discrepant SFRs also tend to have larger
\oiv/\neii\ ratios, suggesting that the radiation field is dominated
by the AGN.  The typical \oiv/\neii\ ratio for AGNs is $\sim3$
\citep[e.g.,][]{stu02,mel08b,per10}, so for sources with
$\oiv/\neii>3$ and large \neii-to-aromatic-feature ratios, it is
likely that the observed \neii\ is dominated by the AGN rather than
star formation.  For sources with large \neii-to-aromatic-feature
ratios but smaller \oiv/\neii\ ratios, the situation is less clear.
However, the amount of \neii\ produced by the AGN is strongly
dependent on the ionization parameter \citep[e.g.,][]{gro06}, and it
is not uncommon for the observed \oiv/\neii\ ratio for AGNs to be as
large as $\sim10$ or smaller than unity
\citep[e.g.,][Figure~\ref{fig:o4ne2pahsfr}]{per10}.  The two sources
with $\oiv/\neii<0.3$ and EW(11.3~$\mu$m)$<0.1$~$\mu$m are the nearby
spiral galaxy M81 (NGC3031) and the radio galaxy Perseus A
(NGC1275)\footnote{Both of these galaxies have strong silicate
  emission, which decreases their EW(11.3~$\mu$m) values because we
  include silicate emission as a continuum component.  For a detailed
  analysis of M81, see \citet{smi10}.}, both of which may exhibit
advection-dominated accretion flows with softer spectral energy
distributions \citep[e.g.,][]{qua99,bal08,mil10}.

Several authors have discussed the question of what fraction of
\neii\ emission is produced by the AGN
\citep[e.g.,][]{stu02,sch06,mel08b,wea10,per10}.  This fraction can be
estimated by assuming a fiducial relationship between \oiv\ and
\neii\ emission for pure AGNs (i.e., with no star formation
contribution to \neii), and then attributing excess \neii\ emission to
star formation.  For example, \citep{stu02} adopted the ratio
$\oiv/\neii=2.7$ for pure AGNs, and \citet{mel08b} determined a
luminosity-dependent relationship between \neii\ and \oiv\ for Seyfert
galaxies with undetected aromatic features, yielding a pure AGN ratio
of $\oiv/\neii=0.9$ for sources with
$L_{\scriptsize{\oiv}}=10^{39}$~erg~s$^{-1}$ and $\oiv/\neii=3.1$ for
sources with $L_{\scriptsize{\oiv}}=10^{42}$~erg~s$^{-1}$.  We compare
aromatic SFRs to \neii\ SFRs estimated based on this latter method in
Figure~\ref{fig:melendez}.  There is rough agreement in the median SFR
between the two methods, but there is significant disagreement on a
source-by-source basis.  In particular, the \citet{mel08b} method
assigns SFR=0 for a large number of RSA Seyferts with clearly detected
aromatic features, and it gives larger SFRs for a number of sources
with smaller \oiv/\neii\ ratios (e.g., NGC3031 and NGC1275, as
discussed above).

Thus, for any individual Seyfert galaxy, it is not straightforward to
determine an accurate SFR based on its \neii\ emission line, primarily
because the AGN contribution to \neii\ varies significantly from
source to source \citep[e.g.,][]{gro06,per10}.  If one assumed that
all of the \neii\ emission were produced by star formation, one would
overestimate SFRs for AGN-dominated sources and underestimate the
scatter in the relationship between AGN and star-forming luminosity
(e.g., see Figure~\ref{fig:lne2_lpah}), thus concluding that this
relationship is stronger than it actually is.  The destruction of
aromatic molecules by the AGN does not appear to have a significant
effect on the 11.3~$\mu$m aromatic feature in the circumnuclear
environment of local Seyfert galaxies \citep{dia10}.  That said, if
such destruction were important, it would mean that the connection
between star-formation rate and black hole accretion rate (a central
result of this paper, see Section~\ref{sec:results}) is actually
stronger than we've presented.  We therefore adopt the 11.3~$\mu$m
aromatic feature as the most robust tracer of the SFR for our sample.
We adopt an uncertainty of 0.2~dex on conversions between the
11.3~$\mu$m aromatic feature strength and IR luminosity based on the
scatter in this ratio for the SINGS sample \citep{smi07a} and an
additional uncertainty of 0.2~dex for conversions between IR
luminosity and SFR \citep{rie09}.  Adding these in quadrature, the
uncertainty on SFRs obtained from equation~\ref{eq:pah} is 0.28~dex.

\section{Results}\label{sec:results}

\subsection{Black Hole Accretion v. Nuclear Star Formation}\label{sec:sfrnuc}

In Figure~\ref{fig:mdotsfr}, we show the relationship between BHAR, as
traced by \oiv, and nuclear SFR, as traced by the 11.3~$\mu$m aromatic
feature.  A strong correlation is apparent: Seyferts with larger BHARs
tend to have larger nuclear SFRs.  We use the linear regression
method\footnote{code available from the IDL Astronomy User's Library
  (linmix\_err.pro), http://idlastro.gsfc.nasa.gov/} outlined by
\citet{kel07} to quantify the relationship between nuclear SFR and
BHAR:
\begin{equation}
SFR (11.3~\mu m, M_{\odot}~\textnormal{yr}^{-1}) = 7.6^{+9.8}_{-3.9} \left(
\frac{\dot{M}_{BH}}{M_{\odot}~\textnormal{yr}^{-1}} \right) ^{0.80^{+0.14}_{-0.12}}
\label{eq:sfrpah}
\end{equation}
The uncertainties on the regression parameters above correspond to the
interval that includes 90\% of the posterior distribution for each
parameter (see Table~2).  The best-fit regression line
and 95\% confidence interval, given the uncertainties in the
regression parameters, are shown as solid and dashed lines in
Figure~\ref{fig:mdotsfr}.  The observed scatter around this
relationship is 0.52~dex (treating BHAR upper limits as detections),
although the posterior median estimate of the intrinsic scatter is
$0.37$~dex (see Table~2), suggesting that much of the
observed scatter may be driven by the measurement errors on SFR and
BHAR.

\begin{figure}[!t]
\begin{center}
\includegraphics[angle=90,scale=.37]{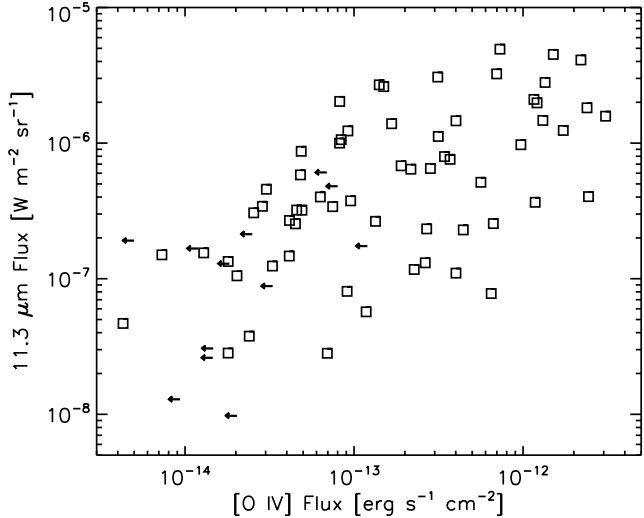}
\caption{The relationship between \oiv\ flux and 11.3~$\mu$m aromatic
  feature flux.  The correlation between these two observed quantities
  is statistically significant, illustrating that the connection
  between the derived physical quantities (nuclear SFR and BHAR) is
  real and not just driven by the distance-squared factor in
  luminosity--luminosity plots.}
\label{fig:o4pah}
\end{center}
\end{figure}

To test whether the relationship between SFR and BHAR could be driven
by the distance dependence inherent in luminosity--luminosity plots,
in Figure~\ref{fig:o4pah} we show the relationship between the
observed quantities, 11.3~$\mu$m aromatic feature and \oiv\ flux.  The
correlation in this flux--flux plot is still statistically significant
\citep[Spearman's $\rho$=0.66, probability of no correlation
  $p<1\times10^{-6}$;][]{iso86,lav92}, confirming the reality of this
relationship.

\begin{figure}[!t]
\begin{center}
\hspace{-1cm}
\includegraphics[angle=90,scale=.37]{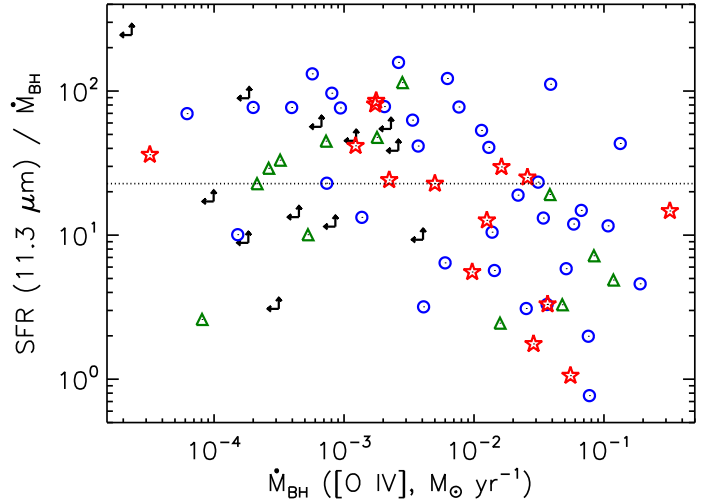}
\caption{The nuclear SFR/BHAR ratio as a function of BHAR.  The median
  ratio SFR/BHAR=23 is shown as a dotted line.  There is mild
  anti-correlation such that sources with large accretion rates tend
  to have smaller SFR/BHAR ratios.  Plot symbols are the same as in
  Figure~\ref{fig:ml}. }
\label{fig:mdotsfr2}
\end{center}
\end{figure}

We investigate the behavior of the SFR/BHAR ratio as a function of
BHAR in Figure~\ref{fig:mdotsfr2}; the median ratio SFR/BHAR=23 is
shown as a dotted line.  A mild anti-correlation exists such that
sources with large accretion rates tend to have smaller SFR/BHAR
ratios (Spearman's $\rho$=-0.48, $p=8\times10^{-5}$).  Thus,
consistent with the sub-linear slope in equation~\ref{eq:sfrpah}, this
indicates that the nuclear SFR does not keep pace with the BHAR
towards high AGN luminosities.  We find no significant difference in
the nuclear SFRs or SFR/BHAR ratios between different Seyfert types
(see Section~\ref{sec:type}).

\subsubsection{Nuclear Star Formation in Physically Matched Apertures}

\begin{figure}[!t]
\begin{center}
\hspace{-1.35cm}
\includegraphics[angle=90,scale=.37]{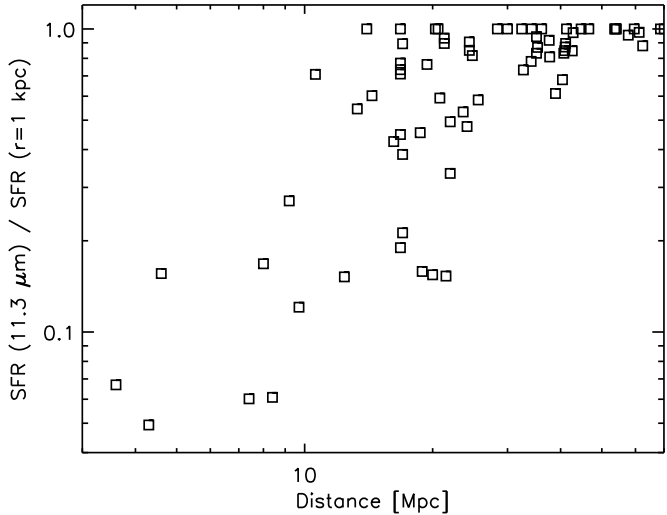}
\caption{The ratio of the the SFR determined from the 11.3~$\mu$m
  feature to the SFR in a $r=1$~kpc aperture, where additional
  contributions from 24~$\mu$m emission outside the spectroscopic
  aperture are included, as a function of galaxy distance.  
  Such contributions are often significant for
  sources with $D<30$~Mpc.}
\label{fig:dist}
\end{center}
\end{figure}

As described in Section~\ref{sec:data}, we also measured 24~$\mu$m
fluxes inside $r=1$~kpc apertures to assess contributions from star
formation that falls outside our IRS extraction aperture.  In
Figure~\ref{fig:dist}, we show that this contribution can be
significant for sources with $D<30$~Mpc.  Since the nearby sources
tend to be have smaller BHARs, including this additional
star-formation contribution leads to a shallower relationship between
nuclear SFR and BHAR.  We show this relationship in
Figure~\ref{fig:nuc}, which is described by the following:
\begin{equation}
SFR (r=1~\textnormal{kpc}, M_{\odot}~\textnormal{yr}^{-1}) = 4.7^{+6.8}_{-2.3} \left(
\frac{\dot{M}_{BH}}{M_{\odot}~\textnormal{yr}^{-1}} \right) ^{0.61^{+0.15}_{-0.11}}
\label{eq:sfrnuc}
\end{equation}
Again, the uncertainties on the values above correspond to the
interval that includes 90\% of the posterior distribution for each
parameter (see Table~2).  The observed scatter around
this relationship is 0.50~dex and the posterior median estimate of the
intrinsic scatter is 0.41~dex.  The bottom panel of
Figure~\ref{fig:nuc} illustrates how the SFR/BHAR ratio tends to
decrease as a function of BHAR.

It is worth noting that our sample does not include galaxies with
larger SFR/BHAR ratios by definition since this is a sample of AGNs
and is limited to sources that have $\oiv/\neii>0.15$, where we can
accurately estimate AGN luminosities (see Section~\ref{sec:data}).
Based on equations~\ref{eq:oiv} and \ref{eq:ne2}, this cut corresponds to
a SFR/BHAR ratio of $\approx350$, which is similar to the maximum
ratio in Figure~\ref{fig:nuc}.

\subsection{Black Hole Accretion v. Extended Star Formation}\label{sec:sfrext}

The connection described above between black hole accretion and
nuclear star formation motivates us to consider whether such a
relationship also exists between black hole activity and star
formation on larger scales in the host galaxy.  We examine this
relationship using extended ($r>1$~kpc) SFRs estimated from MIPS
24~$\mu$m fluxes (see Section~\ref{sec:data}, Table~\ref{tab:data},
and Figure~\ref{fig:extended}) and total SFRs estimated from the
nuclear and extended components (see Figure~\ref{fig:total}).  

\begin{figure}[!t]
\begin{center}
\includegraphics[angle=0,scale=.45]{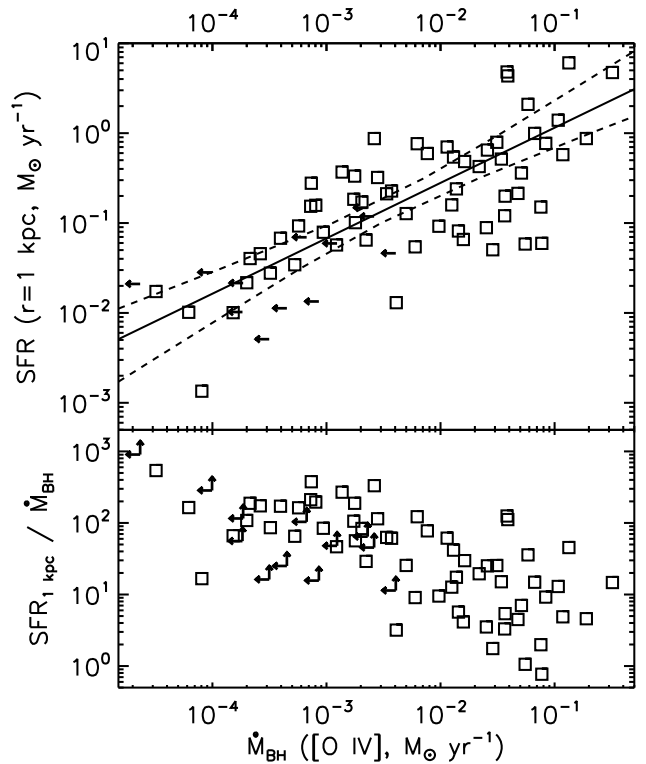}
\caption{The relationship between nuclear SFR (measured on $r=1$~kpc
  scales) and the BHAR.  A strong correlation exists, and the slope of
  the relationship is sub-linear as sources with larger BHARs tend to
  have smaller SFR/BHAR ratios (see equation~\ref{eq:sfrnuc}).  }
\label{fig:nuc}
\end{center}
\end{figure}

\begin{figure}[!t]
\begin{center}
\includegraphics[angle=0,scale=.45]{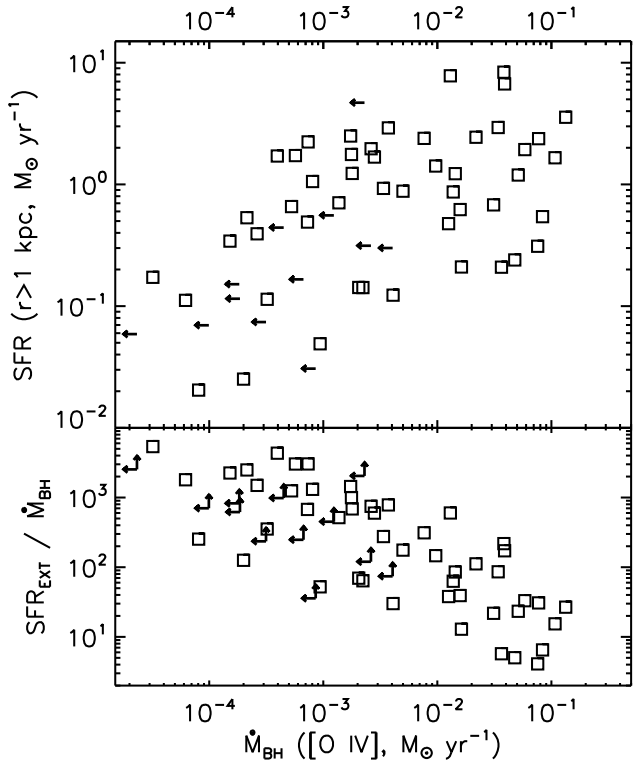}
\caption{The relationship between extended SFR (measured on $r>1$~kpc
  scales) and the BHAR.  Only a mild correlation exists between these
  quantities.}
\label{fig:extended}
\end{center}
\end{figure}

\begin{figure}[!t]
\begin{center}
\includegraphics[angle=0,scale=.45]{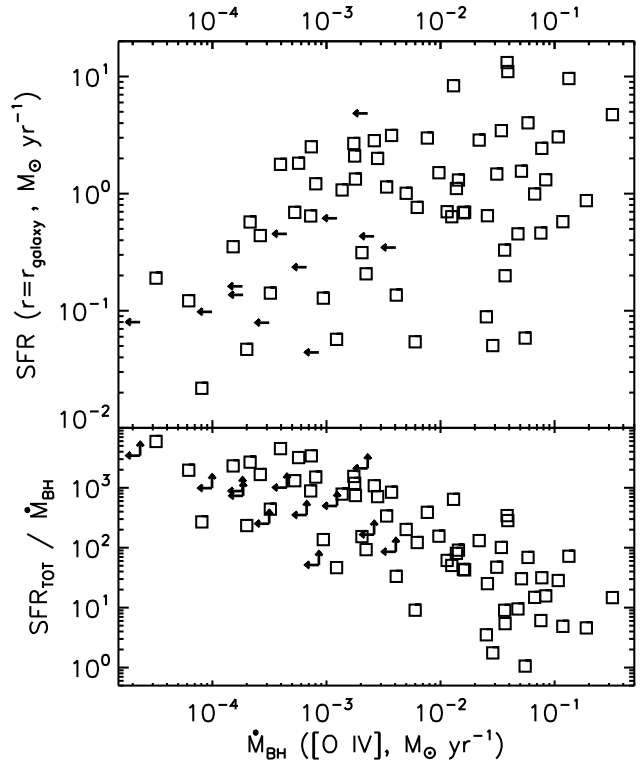}
\caption{The relationship between a galaxy's total SFR and BHAR, which
  exhibits only a weak correlation.}
\label{fig:total}
\end{center}
\end{figure}

There are correlations present in
Figures~\ref{fig:extended}--\ref{fig:total}, but the scatter is
significantly larger than for nuclear SFRs (Figures \ref{fig:mdotsfr}
and \ref{fig:nuc}).  The posterior median estimate of the intrinsic
scatter in the SFR--BHAR relationship is 0.73~dex for $r>1$~kpc SFRs
and 0.86~dex for total SFRs.  Furthermore, the correlation
probabilities for the flux--flux versions of these extended and total
SFR relationships are not highly significant (see
Table~2), illustrating that the connection between the
physical quantities is weak.  In addition, the observed significance
is enhanced by a Malmquist-type bias against more distant galaxies
with lower SFRs, related to the galaxy apparent magnitude limit of the
RSA Seyfert sample.

\begin{figure}[!t]
\begin{center}
\includegraphics[angle=0,scale=.45]{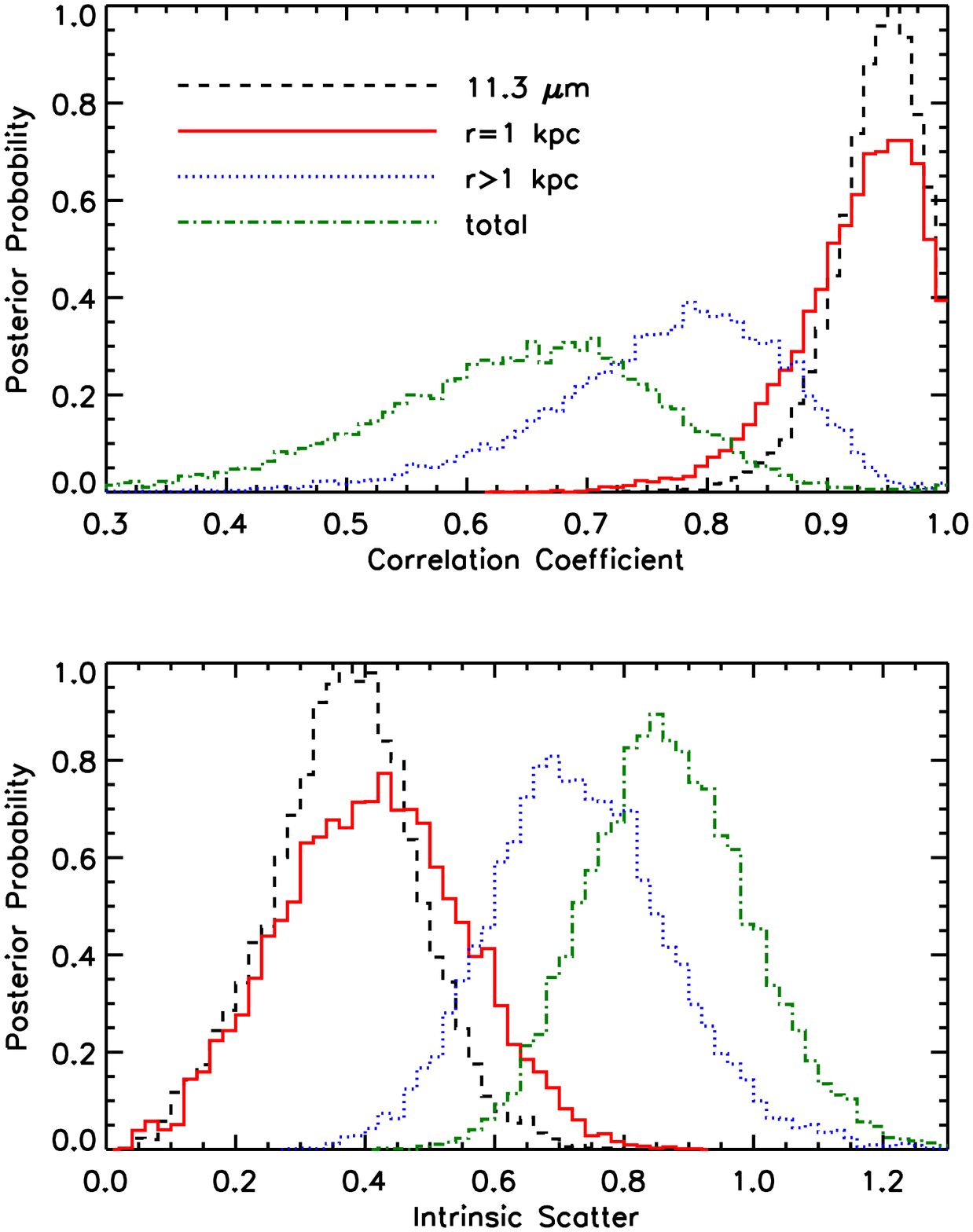}
\caption{The posterior distributions for the correlation coefficient
  and intrinsic scatter in the SFR--BHAR relationships for different
  SFR values.  This figure illustrates that the relationships with
  nuclear SFR exhibit a stronger correlation with smaller scatter
  than the relationships with extended and total star formation.}
\label{fig:corr}
\end{center}
\end{figure}

To illustrate the stronger correlation and smaller scatter associated
with nuclear SFRs, in Figure~\ref{fig:corr} we show the posterior
distributions for the correlation coefficient and intrinsic scatter of
the SFR--BHAR relationship when considering (1) 11.3~$\mu$m aromatic
feature, (2) $r=1$~kpc, (3) $r>1$~kpc, and (4) total galaxy SFRs.
This figure illustrates that while the BHAR correlates reasonably well
with star formation on sub-kpc scales, it is only weakly related to
extended and total star formation activity.

\section{Discussion}

Our results present a picture where the star formation on sub-kpc
scales in AGN host galaxies traces the BHAR in a somewhat sub-linear
fashion, while star formation on larger scales only weakly traces the
BHAR.  Recently, \citet{lut10} argued that host galaxy star formation
only shows a clear dependence on AGN luminosity for high-luminosity
sources ($L_{AGN}>10^{45}$~erg~s$^{-1}$ or
$\dot{M}_{BH}>0.1$~M$_{\odot}$~yr$^{-1}$, see their Figure~6), but our
results show that this relationship persists towards lower AGN
luminosities if one considers only the nuclear component of the host
galaxy.  Given that estimates for samples of AGNs usually provide only
total SFRs (see Section~\ref{sec:comdata}), the observed scaling
between SFR and BHAR may depend on whether the star formation is
dominated by a nuclear or extended component.

\subsection{Comparison with Previous SFR and BHAR Measurements}\label{sec:comdata}

Several authors have estimated SFRs for AGN host galaxies and explored
the relationship with BHAR.  Some find an approximately linear
relationship \citep[$\alpha\geq0.8$, where
  $SFR\propto\dot{M}_{BH}^{\alpha}$, e.g.,][]{sat05,net09,shi09}
consistent with our results for nuclear SFRs traced by the 11.3~$\mu$m
aromatic feature (equation~\ref{eq:sfrpah}, Figure~\ref{fig:mdotsfr}),
while others finding a much shallower relationship
\citep[$\alpha\leq0.5$, e.g.,][]{hao05,sil09,bon11}, more consistent
with our results for extended and total SFRs
(Figures~\ref{fig:extended}--\ref{fig:total}).

Among studies that have used the aromatic features to estimate SFRs,
\citet{net09} compiled Spitzer measurements for 28 $z\sim0.1$ QSOs
\citep{net07} and 12 $z\sim2$ QSOs \citep{lut08} to complement their
own study of a large sample of type 2 Seyferts and LINERs from the
Sloan Digital Sky Survey (SDSS).  The AGN luminosities for the SDSS
sources were estimated using the $\oiii\lambda5007$ and
$\oi\lambda6300$ lines, and SFRs were estimated using the method of
\citet{bri04}.  They found a relationship of the form
$SFR\propto\dot{M}_{BH}^{0.8}$ with a SFR/BHAR ratio $\sim30$ for
$\dot{M}_{BH}=0.1$~M$_{\odot}$~yr$^{-1}$.  Similarly, \citet{shi09}
considered the relationship between 5--6~$\mu$m continuum luminosity
and aromatic-feature luminosity for a sample of 89 PG quasars at
$z<0.5$ \citep{shi07} and 57 SDSS quasars at $z\sim1$ and found a
relationship of the form $SFR\propto\dot{M}_{BH}^{0.97\pm0.08}$ with a
SFR/BHAR ratio $\sim10$ for sources with
$\dot{M}_{BH}=1$~M$_{\odot}$~yr$^{-1}$.  These relationships are
consistent with our results for aromatic-based SFRs
(Section~\ref{sec:sfrnuc}), although they extend to higher
luminosities.

A number of studies have also used measurements of far-IR luminosities
of AGNs to estimate host galaxy SFRs.  There are uncertainties
regarding poorly sampled IR spectral energy distributions and
contributions from dust heated by the AGN
\citep[e.g.,][]{sch06,shi07}, but these estimates are nonetheless
instructive.  For example, \citet{sat05} expanded on the work of
\citet{dud05} and compiled AGN and far-IR luminosities for a sample
including 86 Seyferts and quasars, for which they found a relationship
of the form $SFR\propto\dot{M}_{BH}^{0.89}$ with an SFR/BHAR ratio
$\sim60$ at $\dot{M}_{BH}=0.1$~M$_{\odot}$~yr$^{-1}$.  This is similar
to our Equation~\ref{eq:sfrpah}, but with a somewhat higher SFR
normalization.  On the other hand, a shallower relationship
($\alpha=0.29$) with a much larger SFR normalization
(SFR/BHAR~$\sim3000$ at $\dot{M}_{BH}=0.1$~M$_{\odot}$~yr$^{-1}$) was
found by \citet{hao05} for a sample of 31 AGN-ULIRGs at $z<0.5$, where
excess emission at 60~$\mu$m was attributed to star formation.  More
recently, such a shallow slope ($\alpha\leq0.5$) has also been found
by several studies of optically selected quasars
\citep[e.g.,][]{ser09,hat10,bon11} in fields that have Spitzer or
Herschel \citep{pil10} coverage.

Several studies of X-ray selected AGNs have also found a shallow
SFR--BHAR relationship.  \citet{sil09} used the $\oii\lambda3727$ line
to estimate SFRs for a sample of COSMOS \citep{sco07} X-ray AGNs and
found $SFR\propto\dot{M}_{BH}^{0.28\pm0.22}$ with a SFR/BHAR ratio
$\sim50$ at $\dot{M}_{BH}=0.1$~M$_{\odot}$~yr$^{-1}$.  \citet{atl11}
considered a sample of X-ray and IR-selected AGNs in galaxy clusters,
with SFRs estimated from spectral decompositions of mid-IR data
\citep{ass10}, and found $SFR\propto\dot{M}_{BH}^{0.46\pm0.06}$.
While these sources reside in denser environments, their AGN
luminosities and total SFRs fall in the range of RSA Seyferts.
\citet{lut10} and \citet{sha10} measured 870~$\mu$m and
100--160~$\mu$m emission, respectively, for samples of Chandra
X-ray-selected AGNs \citep{ale03,leh05,toz06}, and compiled 60~$\mu$m
measurements for local AGNs detected by Swift-BAT \citep{cus10}.  They
also find a shallow SFR--BHAR slope ($\alpha\sim0.4$), but argue that
star formation and black hole growth are more closely linked at higher
AGN luminosities.  Recently, \citet{mul11} found no correlation
between X-ray luminosity and far-IR luminosity for moderate luminosity
X-ray sources ($L_{X}=10^{42}$--$10^{44}$~erg~s$^{-1}$) up to
$z\approx3$, consistent with the weak relationship we find for total
SFRs.

\subsection{Comparison with Model Predictions}

A number of authors have made theoretical predictions for the behavior
of the SFR and the BHAR during the AGN phase.  These models differ
primarily in their predictions for the nuclear SFR/BHAR ratio, ranging
from $\sim1$ \citep[e.g.,][]{kaw08} to $\sim10^3$
\citep[e.g.,][]{tho05} for AGNs with
$\dot{M}_{BH}\sim0.1$~M$_{\odot}$~yr$^{-1}$.  

For example, in the galaxy merger models of \citet{dim05} that produce
a final black hole mass $\sim4\times10^{7}$~M$_{\odot}$, which is
appropriate for our sample (see Figure~\ref{fig:ml}), the SFR/BHAR
ratio reaches $\sim200$ at the peak of star-formation activity as the
galaxies coalesce and drops to $\sim5$ at the end of the bright AGN
phase ($\dot{M}_{BH}>0.1$~M$_{\odot}$~yr$^{-1}$).  Although the RSA
Seyferts show little evidence for merger activity, our results for
nuclear SFRs are broadly consistent with these values.  For lower
accretion rates
($\dot{M}_{BH}=10^{-3}$--$10^{-2}$~M$_{\odot}$~yr$^{-1}$)
characteristic of earlier merger phases in the \citet{dim05} model,
the SFR/BHAR ratio falls in the 500--1000 range, consistent with our
results for total SFRs (see Figure~\ref{fig:total}).  

The starburst disk models of \citet{tho05} that produce
$\dot{M}_{BH}\sim0.1$~M$_{\odot}$~yr$^{-1}$ suggest that most of the
gas being supplied at an outer radius $R_{out}=200$~pc will be
consumed by a starburst near that outer radius, resulting in
SFR/BHAR$\sim10^3$.  While inconsistent with our measurements for
local Seyfert galaxies, this model is perhaps consistent with the
findings of \citet{hao05} for AGN-ULIRGs.  \citet{ball08} used a
scaled-down version of the \citet{tho05} model to study less powerful
starbursts around AGNs. They found SFR/BHAR$\sim200$ inside $r=100$~pc
for a fiducial model with $\dot{M}_{BH}\sim0.3$~M$_{\odot}$~yr$^{-1}$,
which is more star formation than we observe in $r=1$~kpc apertures
(see Figure~\ref{fig:nuc}).  They also predict that the SFR/BHAR ratio
should increase toward lower AGN luminosities due to competition for
gas between between star formation and black hole accretion.  This
model assumes a constant gas supply, and could perhaps be reconciled
with our observations if the gas supply were depleted in such a way
that the SFR decreased more quickly than the BHAR.

\citet{esc07} studied the evolution of the central kpc of a massive
nuclear disk with a central black hole and find that the SFR and BHAR
trace each other during the primary growth phase
($SFR\propto\dot{M}_{BH}$) with SFR/BHAR ratios in the 3--50 range.
This is consistent with our results regarding nuclear SFRs for sources
with $\dot{M}_{BH}>0.01$~M$_{\odot}$~yr$^{-1}$ (see
Figure~\ref{fig:nuc}).  Focusing on smaller scales, \citet{kaw08}
studied a 100~pc circumnuclear disk with accretion being driven by
turbulence from supernovae.  They found a correlation between SFR and
BHAR only for sources with $\dot{M}_{BH}>0.01$~M$_{\odot}$~yr$^{-1}$,
and they predict that the SFR/BHAR ratio should increase with AGN
luminosity, reaching a maximum ratio $\sim2$.  This model predicts
less circumnuclear star formation than we observe in $r=1$~kpc
apertures, and can only be reconciled with our observations if there
is $10\times$ more star formation on 100~pc$<r<1$~kpc scales than on
$r<100$~pc scales.

\citet{hop10} went a step further and made predictions for the
relationship between SFR and BHAR as a function of radius.  They used
results from large-scale (100~kpc to 100~pc) simulations of galaxy
mergers and barred galaxy disks, and then re-simulated the central kpc
(1~kpc to 10~pc) and the central 10~pc (10~pc to 0.1~pc) at higher
spatial resolution.  They find SFR/BHAR ratios ranging from $\sim10$
for $R<100$~pc to $\sim30$ for $R<1$~kpc, and $\sim300$ for the whole
galaxy, which is consistent with our results for nuclear and total
SFRs.  They also find $SFR\propto\dot{M}_{BH}$ on the smallest scales
($R<10$~pc) and $SFR\propto\dot{M}_{BH}^{0.7}$ on larger scales, which
is in general agreement with the scalings we find for nuclear star
formation.

\subsection{Implications for Black Hole and Bulge Growth}

It is worthwhile to consider our measurements of nuclear SFRs and
BHARs in terms of the observed scaling between galaxy bulge and black
hole masses \citep[e.g.,][]{mar03,har04}.  \citet{hec04} analyzed a
sample of bulge-dominated galaxies from the Sloan Digital Sky Survey
and found that the integrated black hole growth (from galaxies hosting
AGNs) and star formation (from all galaxies) corresponded to
SFR/BHAR~$\sim10^3$, consistent with the observed bulge/black hole
mass ratio.  \citet{fu10} performed a similar exercise for a sample of
luminous infrared galaxies and found similar results for the
integrated population.  For our sample, we find a median
SFR/BHAR~$=36$ on scales of $r=1$~kpc (the integrated total
corresponds to SFR/BHAR~$=23$), which is factor of $\sim20$ below the
median bulge/black hole mass ratio \citep{har04}.  This implies an AGN
duty cycle of $\sim5\%$, which is within a factor of two of the
Seyfert fraction in the RSA galaxy sample \citep[$\sim10\%$,][]{ho97}.

There are several caveats associated with the above estimate,
including the fact we excluded the most star-formation dominated
Seyferts (see Section~\ref{sec:data}) and that the our nuclear
apertures have not been customized to match the bulge of each galaxy.
A systematic bulge/disk decomposition is beyond the scope of this
work, but $r=1$~kpc corresponds roughly to the effective radius of a
$10^{10}~M_{\Sun}$ bulge \citep[][]{mar03,she03,gra08b}, meaning that
our nuclear apertures encompass half the light of such a bulge (and
less light for more massive bulges).  Accounting for these effects, as
well as potential bulge growth through dynamical process that relax
the orbits of pre-existing stars, would tend to increase our estimate
of the AGN duty cycle.  We have also assumed that local Seyfert
galaxies obey the black hole--bulge scaling relations defined
primarily by early-type galaxies with classical bulges, which may or
may not be the case \citep[e.g.,][]{gra08a,hu08,gre10,kor11}.

We now consider our results in the context of AGN fueling mechanisms.
A number of morphological studies of AGN host galaxies at $z<1$ have
argued that the fueling for most systems is not merger driven
\citep[e.g.,][]{gro05,pie07,gab09,cis11}.  While large-scale bars can
also provide the necessary gravitational torques to drive gas down to
$\sim100$~pc, there is not strong evidence that Seyfert galaxies
exhibit a larger bar fraction than do star-forming galaxies
\citep[e.g.,][]{ho97b,mul97,hun99}.  However, both Seyfert and
star-forming activity appear to be associated with a higher incidence
of bars than found in quiescent galaxies \citep{lau04,hao09}.  We have
demonstrated that the BHAR is correlated with the amount of gas on
sub-kpc scales \citep[adopting the $SFR\propto\Sigma_{gas}$ assumption
  of the Schmidt-Kennicutt relationship,][]{sch59,ken98b}, but such
gas still needs to shed most of its angular momentum to reach the
black hole.  Future studies with ALMA of the spatial distribution and
kinematics of molecular gas down to $\sim$pc scales in local Seyfert
galaxies will probe the nature of this fueling more directly.  We note
that \citet{dav07} and \citet{wil10} presented evidence for a time
delay between the onset of star formation and AGN activity, which can
be explained if the black hole is being fed by outflows from
intermediate-age stars.  Our observations are not sensitive to
detecting such a $\sim$100~Myr time delay because the aromatic
features will continue to be excited by UV photons from longer-lived B
stars \citep[e.g.,][]{pee04,diaz10}.  The required loss of angular
momentum could also be explained by dynamical instabilities in
self-gravitating disks \citep[e.g.,][]{hop10}.

\subsection{Behavior as a Function of Seyfert Type}\label{sec:type}

There have been suggestions from theoretical
\citep[e.g.,][]{wad02,bal06} and observational work
\citep[e.g.,][]{mai95b,mou02,buc06,deo07,mel08b} that star formation
is enhanced in obscured (i.e., type 2) AGNs.  This would be
inconsistent with the standard unified model
\citep[e.g.,][]{ant93,urr95} where differences between obscured and
unobscured AGNs are attributed to our viewing angle towards a central
obscuring torus, and would suggest that the obscuring material in type
2 AGNs is related to star-formation activity in the host galaxy.
However, we find that the distributions of nuclear SFRs, extended
SFRs, total SFRs, and SFR/BHAR ratios for our sample do not exhibit
any statistically significant differences between type 1 and type 2
Seyferts.  While this result does not definitively rule out any
difference between the star-forming properties of type 1 and type 2
AGNs, it does imply that such differences are not dramatic.

We note that results on star-formation activity as a function of
Seyfert type likely depend on sample selection and the method used to
measure the SFR.  A comprehensive analysis of these factors is beyond
the scope of this work, but their effect is apparent in previous
published results.  For example, \citet{mai95b} studied a sample
including 51 Seyferts from the CfA sample \citep{huc92}, 59 from the
RSA sample, and 84 from the 12~$\mu$m sample \citep{rus93}.  By
comparing ground-based 10~$\mu$m observations with IRAS 12--100~$\mu$m
fluxes, they found that the extended IR emission in type 2 Seyferts
tends to have a much redder color than that in type 1s, consistent
with enhanced star formation in the type 2 galaxies.  Similarly,
\citet{mou02} studied a sample of 50 RSA and 37 CfA Seyferts and found
that Seyfert 2s had larger infrared/B-band and larger
100~$\mu$m/60~$\mu$m flux ratios, implying that they were more often
starburst or host-galaxy dominated.  On the other hand, \citet{ima04}
studied a sample of 24 CfA and 33 12~$\mu$m Seyferts and found no
significant difference in 3.3~$\mu$m aromatic feature luminosities
between type 1 and type 2 Seyferts.  More recently, \citet{per10}
found a significant enhancement in the fraction of type 2 Seyferts
that exhibit excess \neiii\ emission associated with star formation
for a heterogeneous sample of 201 Seyferts, but found no significant
difference when considering more complete samples of 70 RSA and 97
12~$\mu$m Seyferts.

\section{Conclusions}

We have measured BHARs based on the \oiv\ emission line and nuclear,
extended, and total SFRs based on the 11.3~$\mu$m aromatic feature and
extended 24~$\mu$m continuum emission for a complete sample of Seyfert
galaxies.  We find a strong correlation between the nuclear star
formation and black hole accretion, where the nuclear star formation
is traced by the 11.3~$\mu$m aromatic feature and 24~$\mu$m emission
external to the central PSF but within $r=1$~kpc (see
Figures~\ref{fig:mdotsfr} and \ref{fig:nuc}).  On the other hand, the
extended ($r>1$~kpc) and total SFRs are only weakly correlated with
the BHAR (see Figure~\ref{fig:extended}--\ref{fig:corr} and
Table~2).  This suggests a connection between gas on
sub-kpc scales that is forming stars and the gas on sub-pc scales that
is accreting onto the black hole.  While the physical processes that
drive this relationship (e.g., mass loss from evolved stars, angular
momentum loss from gravitational instabilities) are not clearly
identified by these data, the connection is apparently unrelated to
external processes on kpc scales in the host galaxy.

\acknowledgments

We thank the anonymous referee for helpful suggestions that have
improved the paper.  We acknowledge useful discussions with colleagues
at all five Center for Galaxy Evolution campuses (UCI, UCLA, UCR,
UCSB, UCSD), where this work was presented prior to submission.  We
also acknowledge constructive feedback from David Ballantyne, Jill
Bechtold, Xiaohui Fan, Alister Graham, Richard Green, Marcio
Mel{\'e}ndez, John Moustakas, Greg Novak, Feryal Ozel, and Miguel
Pereira-Santaella.  AMD acknowledges support from the Southern
California Center for Galaxy Evolution, a multi-campus research
program funded by the University of California Office of Research.
This work was also partially supported by contract 1255094 from
Caltech/JPL to the University of Arizona.

{\it Facilities:} \facility{Spitzer}

\clearpage

\begin{landscape}
\LongTables
\begin{deluxetable}{lcccccrlr}
\tabletypesize{\scriptsize}
\tablecaption{Measurements and Derived Parameters\label{tab:data}}
\tablewidth{0pt}
\tablehead{
\colhead{NAME} & \colhead{11.3~$\mu$m} & \colhead{\neii} & \colhead{aperture} & \colhead{SFR} & \colhead{SFR} & \colhead{$\dot{M}_{BH}$} & \colhead{M$_{BH}$} & \colhead {L/L$_{Edd}$} \\
& \colhead{intensity, EW, SFR} & \colhead{intensity, SFR} & & \colhead{$r=1$~kpc} & \colhead{$r>1$~kpc} & & \colhead{value, method, ref.} & \\
\colhead{(1)} & \colhead{(2)} & \colhead{(3)} & \colhead{(4)} & \colhead{(5)} & \colhead{(6)} & \colhead{(7)} & \colhead{(8)} & \colhead{(9)} }
\startdata
   IC2560  &  5.14$\pm$0.20e$-07$, 0.10, 3.0e$-$01  &  1.58$\pm$0.03e$-07$, 9.7e$-$01  &  $0.71\times1.42$  &   3.6e$-$01  &   1.2e+00  &   5.1e$-$02  &  \nodata   &  \nodata   \\
   IC3639  &  1.39$\pm$0.05e$-06$, 0.20, 6.1e$-$01  &  3.11$\pm$0.03e$-07$, 1.4e+00  &  $0.62\times1.23$  &   7.0e$-$01  &   \nodata  &   1.1e$-$02  &   6.8e+06, dis, 1  &   7.2e$-$02  \\
IRAS11215  &  5.70$\pm$1.40e$-08$, 0.02, 7.8e$-$02  &  1.67$\pm$0.02e$-08$, 2.4e$-$01  &  $1.09\times2.18$  &   8.9e$-$02  &   \nodata  &   2.5e$-$02  &  \nodata   &  \nodata   \\
   MRK509  &  6.50$\pm$0.14e$-07$, 0.10, 4.7e+00  &  1.05$\pm$0.02e$-07$, 8.0e+00  &  $2.51\times5.02$  &   4.7e+00  &   \nodata  &   3.2e$-$01  &   1.4e+08, rev, 2  &   9.6e$-$02  \\
   NGC788  &  1.17$\pm$0.15e$-07$, 0.04, 1.2e$-$01  &  5.16$\pm$0.17e$-08$, 5.6e$-$01  &  $0.94\times1.89$  &   1.2e$-$01  &   2.1e$-$01  &   3.6e$-$02  &   3.2e+07, dis, 1  &   4.8e$-$02  \\
  NGC1058  &  1.91$\pm$0.06e$-07$, 1.68, 5.7e$-$03  &  8.94$\pm$1.10e$-09$, 2.8e$-$03  &  $0.16\times0.32$  &   2.1e$-$02  &   5.9e$-$02  &  $<$ 2.3e$-$05  &  \nodata   &  \nodata   \\
  NGC1097  &  1.50$\pm$0.03e$-06$, 1.07, 1.4e$-$01  &  1.44$\pm$0.03e$-07$, 1.5e$-$01  &  $0.29\times0.58$  &   2.3e+00  &   2.3e+00  &  \nodata  &  \nodata   &  \nodata   \\
  NGC1241  &  5.83$\pm$1.01e$-07$, 0.83, 5.9e$-$01  &  7.88$\pm$0.57e$-08$, 8.5e$-$01  &  $0.94\times1.88$  &   5.9e$-$01  &   2.4e+00  &   7.7e$-$03  &   2.9e+07, dis, 1  &   1.1e$-$02  \\
  NGC1275  &  3.75$\pm$0.16e$-07$, 0.03, 6.5e$-$01  &  3.17$\pm$0.03e$-07$, 5.8e+00  &  $1.22\times2.45$  &   6.5e$-$01  &   \nodata  &   2.6e$-$02  &   3.8e+08, dis, 3  &   2.9e$-$03  \\
  NGC1358  &  8.07$\pm$1.81e$-08$, 0.27, 8.2e$-$02  &  2.79$\pm$0.31e$-08$, 3.0e$-$01  &  $0.94\times1.87$  &   8.2e$-$02  &   1.2e+00  &   1.4e$-$02  &   7.6e+07, dis, 1  &   8.1e$-$03  \\
  NGC1365  &  4.51$\pm$0.05e$-06$, 0.34, 7.3e$-$01  &  6.07$\pm$0.06e$-07$, 1.0e+00  &  $0.38\times0.75$  &   4.8e+00  &   8.4e+00  &   3.8e$-$02  &  \nodata   &  \nodata   \\
  NGC1386  &  9.73$\pm$0.34e$-07$, 0.13, 3.8e$-$02  &  1.18$\pm$0.04e$-07$, 4.9e$-$02  &  $0.19\times0.37$  &   5.4e$-$02  &   \nodata  &   6.0e$-$03  &   1.7e+07, dis, 1  &   1.5e$-$02  \\
  NGC1433  &  6.08$\pm$0.15e$-07$, 1.40, 3.8e$-$02  &  4.96$\pm$0.21e$-08$, 3.3e$-$02  &  $0.23\times0.46$  &   6.9e$-$02  &   1.7e$-$01  &  $<$ 6.7e$-$04  &  \nodata   &  \nodata   \\
  NGC1566  &  1.06$\pm$0.01e$-06$, 0.56, 1.4e$-$01  &  7.46$\pm$0.20e$-08$, 1.0e$-$01  &  $0.34\times0.68$  &   1.8e$-$01  &   2.5e+00  &   1.7e$-$03  &  \nodata   &  \nodata   \\
  NGC1667  &  4.01$\pm$0.38e$-07$, 1.16, 5.3e$-$01  &  4.97$\pm$0.28e$-08$, 6.9e$-$01  &  $1.07\times2.14$  &   5.4e$-$01  &   7.8e+00  &   1.3e$-$02  &   7.6e+07, dis, 1  &   7.3e$-$03  \\
  NGC2273  &  2.69$\pm$0.03e$-06$, 0.50, 7.6e$-$01  &  3.20$\pm$0.03e$-07$, 9.6e$-$01  &  $0.50\times0.99$  &   7.6e$-$01  &   \nodata  &   6.2e$-$03  &   2.0e+07, dis, 1  &   1.3e$-$02  \\
  NGC2639  &  1.34$\pm$0.05e$-07$, 0.53, 8.6e$-$02  &  7.55$\pm$0.09e$-08$, 5.1e$-$01  &  $0.74\times1.49$  &   1.0e$-$01  &   1.2e+00  &   1.8e$-$03  &   8.7e+07, dis, 3  &   8.8e$-$04  \\
  NGC2655  &  4.81$\pm$0.19e$-07$, 0.47, 1.0e$-$01  &  7.10$\pm$0.35e$-08$, 1.6e$-$01  &  $0.43\times0.85$  &   1.2e$-$01  &   3.1e$-$01  &  $<$ 2.6e$-$03  &   5.5e+07, dis, 3  &  $<$ 2.0e$-$03  \\
  NGC2685  &  4.68$\pm$0.29e$-08$, 0.24, 4.3e$-$03  &  6.47$\pm$0.56e$-09$, 6.3e$-$03  &  $0.28\times0.57$  &   1.0e$-$02  &   1.1e$-$01  &   6.2e$-$05  &   6.4e+06, dis, 3  &   4.1e$-$04  \\
  NGC2992  &  1.47$\pm$0.02e$-06$, 0.28, 6.0e$-$01  &  3.26$\pm$0.03e$-07$, 1.4e+00  &  $0.60\times1.19$  &   7.7e$-$01  &   5.4e$-$01  &   8.4e$-$02  &  \nodata   &  \nodata   \\
  NGC3031  &  2.54$\pm$0.16e$-07$, 0.06, 1.2e$-$03  &  1.79$\pm$0.03e$-07$, 8.6e$-$03  &  $0.06\times0.13$  &   1.7e$-$02  &   1.7e$-$01  &   3.2e$-$05  &   8.0e+07, gas, 4  &   1.7e$-$05  \\
  NGC3079  &  2.12$\pm$0.03e$-05$, 9.70, 3.1e+00  &  1.17$\pm$0.01e$-06$, 1.8e+00  &  $0.36\times0.71$  &   3.4e+00  &   2.1e+00  &  \nodata  &   \nodata  &  \nodata  \\
  NGC3081  &  3.66$\pm$0.12e$-07$, 0.06, 1.5e$-$01  &  9.60$\pm$0.10e$-08$, 4.2e$-$01  &  $0.60\times1.19$  &   1.5e$-$01  &   3.1e$-$01  &   7.6e$-$02  &   2.3e+07, dis, 1  &   1.4e$-$01  \\
  NGC3147  &  2.13$\pm$0.19e$-07$, 0.24, 1.3e$-$01  &  3.10$\pm$0.30e$-08$, 1.9e$-$01  &  $0.71\times1.43$  &   1.5e$-$01  &   4.7e+00  &  $<$ 2.3e$-$03  &   2.0e+08, dis, 3  &  $<$ 5.0e$-$04  \\
  NGC3185  &  1.00$\pm$0.04e$-06$, 1.60, 1.6e$-$01  &  7.56$\pm$0.36e$-08$, 1.3e$-$01  &  $0.37\times0.74$  &   1.7e$-$01  &   1.4e$-$01  &   2.0e$-$03  &   3.3e+06, dis, 3  &   2.7e$-$02  \\
  NGC3227  &  3.24$\pm$0.03e$-06$, 0.46, 4.8e$-$01  &  5.15$\pm$0.05e$-07$, 8.1e$-$01  &  $0.36\times0.72$  &   4.8e$-$01  &   2.1e$-$01  &   1.6e$-$02  &  4.2e+07, rev, 5   & 1.6e$-$02   \\
  NGC3254  &  3.06$\pm$0.38e$-08$, 0.46, 6.0e$-$03  &  2.91$\pm$0.60e$-09$, 6.0e$-$03  &  $0.41\times0.82$  &   1.1e$-$02  &   4.4e$-$01  &  $<$ 4.5e$-$04  &   1.6e+07, dis, 3  &  $<$ 1.2e$-$03  \\
  NGC3281  &  1.24$\pm$0.25e$-06$, 0.10, 8.7e$-$01  &  1.77$\pm$0.03e$-07$, 1.3e+00  &  $0.78\times1.56$  &   8.7e$-$01  &   \nodata  &   1.9e$-$01  &   1.9e+07, dis, 1  &   4.2e$-$01  \\
  NGC3486  &  8.82$\pm$1.70e$-08$, 1.24, 1.7e$-$03  &  5.83$\pm$3.20e$-09$, 1.2e$-$03  &  $0.13\times0.26$  &   2.8e$-$02  &   7.0e$-$02  &  $<$ 9.9e$-$05  &   1.5e+06, dis, 3  &  $<$ 2.9e$-$03  \\
  NGC3516  &  2.29$\pm$0.26e$-07$, 0.04, 1.2e$-$01  &  4.54$\pm$0.39e$-08$, 2.5e$-$01  &  $0.68\times1.36$  &   2.0e$-$01  &   \nodata  &   3.7e$-$02  &   4.3e+07, rev, 2  &   3.7e$-$02  \\
  NGC3735  &  7.59$\pm$0.30e$-07$, 0.31, 4.5e$-$01  &  5.84$\pm$0.37e$-08$, 3.6e$-$01  &  $0.72\times1.43$  &   5.1e$-$01  &   2.9e+00  &   3.4e$-$02  &   3.3e+07, dis, 3  &   4.5e$-$02  \\
  NGC3783  &  1.10$\pm$0.20e$-07$, 0.01, 5.0e$-$02  &  1.44$\pm$0.21e$-07$, 7.0e$-$01  &  $0.63\times1.26$  &   5.0e$-$02  &   \nodata  &   2.9e$-$02  &   3.0e+07, rev, 2  &   4.1e$-$02  \\
  NGC3941  &  1.29$\pm$0.35e$-08$, 0.05, 1.6e$-$03  &  6.25$\pm$0.61e$-09$, 8.3e$-$03  &  $0.33\times0.66$  &   1.0e$-$02  &   1.5e$-$01  &  $<$ 1.8e$-$04  &   2.6e+07, dis, 3  &  $<$ 3.0e$-$04  \\
  NGC3976  &  1.50$\pm$0.25e$-07$, 0.71, 7.5e$-$02  &  1.34$\pm$0.30e$-08$, 7.1e$-$02  &  $0.66\times1.32$  &   9.3e$-$02  &   1.7e+00  &   5.7e$-$04  &   1.1e+08, dis, 3  &   2.2e$-$04  \\
  NGC3982  &  3.21$\pm$0.19e$-07$, 0.67, 3.3e$-$02  &  5.59$\pm$0.27e$-08$, 6.0e$-$02  &  $0.30\times0.59$  &   1.5e$-$01  &   4.9e$-$01  &   7.3e$-$04  &   1.2e+06, dis, 1  &   2.5e$-$02  \\
  NGC4051  &  1.12$\pm$0.03e$-06$, 0.12, 1.1e$-$01  &  1.04$\pm$0.06e$-07$, 1.1e$-$01  &  $0.30\times0.59$  &   1.3e$-$01  &   8.8e$-$01  &   5.0e$-$03  &   1.7e+06, rev, 5  &   1.2e$-$01  \\
  NGC4138  &  1.05$\pm$0.03e$-07$, 0.18, 1.1e$-$02  &  1.71$\pm$0.05e$-08$, 1.8e$-$02  &  $0.30\times0.59$  &   2.8e$-$02  &   1.1e$-$01  &   3.2e$-$04  &   1.8e+07, dis, 3  &   7.7e$-$04  \\
  NGC4151  &  4.03$\pm$1.72e$-07$, 0.02, 5.8e$-$02  &  2.46$\pm$0.07e$-07$, 3.8e$-$01  &  $0.35\times0.71$  &   5.8e$-$02  &   \nodata  &   5.5e$-$02  &   4.6e+07, rev, 6  &   5.2e$-$02  \\
  NGC4235  &  1.24$\pm$0.15e$-07$, 0.16, 5.4e$-$02  &  2.99$\pm$0.29e$-08$, 1.4e$-$01  &  $0.61\times1.23$  &   6.5e$-$02  &   1.4e$-$01  &   2.2e$-$03  &   4.4e+07, dis, 3  &   2.2e$-$03  \\
  NGC4258  &  3.40$\pm$0.03e$-07$, 0.15, 7.7e$-$03  &  7.47$\pm$0.07e$-08$, 1.8e$-$02  &  $0.14\times0.28$  &   4.6e$-$02  &   3.9e$-$01  &   2.6e$-$04  &   3.8e+07, mas, 7  &   3.0e$-$04  \\
  NGC4378  &  1.29$\pm$0.04e$-07$, 0.62, 5.6e$-$02  &  5.96$\pm$0.73e$-09$, 2.7e$-$02  &  $0.61\times1.23$  &   5.9e$-$02  &   5.6e$-$01  &  $<$ 1.2e$-$03  &   1.1e+08, dis, 3  &  $<$ 4.6e$-$04  \\
  NGC4388  &  1.58$\pm$0.03e$-06$, 0.23, 1.6e$-$01  &  4.98$\pm$0.05e$-07$, 5.2e$-$01  &  $0.29\times0.59$  &   2.1e$-$01  &   2.4e$-$01  &   4.8e$-$02  &   1.7e+07, dis, 1  &   1.2e$-$01  \\
  NGC4395  &  2.82$\pm$0.31e$-08$, 0.18, 2.1e$-$04  &  2.64$\pm$0.04e$-08$, 2.1e$-$03  &  $0.08\times0.16$  &   1.3e$-$03  &   2.0e$-$02  &   8.1e$-$05  &   3.6e+05, rev, 2  &   9.6e$-$03  \\
  NGC4477  &  1.55$\pm$0.04e$-07$, 0.62, 1.5e$-$02  &  1.96$\pm$0.07e$-08$, 2.0e$-$02  &  $0.29\times0.59$  &   2.2e$-$02  &   2.5e$-$02  &   2.0e$-$04  &   8.2e+07, dis, 3  &   1.0e$-$04  \\
  NGC4501  &  3.06$\pm$0.03e$-07$, 0.87, 3.0e$-$02  &  3.94$\pm$0.05e$-08$, 4.1e$-$02  &  $0.29\times0.59$  &   6.8e$-$02  &   1.7e+00  &   4.0e$-$04  &   7.2e+07, dis, 1  &   2.3e$-$04  \\
  NGC4507  &  7.95$\pm$0.18e$-07$, 0.06, 9.9e$-$01  &  2.81$\pm$0.03e$-07$, 3.7e+00  &  $1.04\times2.08$  &   9.9e$-$01  &   \nodata  &   6.7e$-$02  &   3.8e+07, dis, 1  &   7.5e$-$02  \\
  NGC4565  &  1.47$\pm$0.19e$-07$, 0.37, 4.9e$-$03  &  1.99$\pm$0.29e$-08$, 6.9e$-$03  &  $0.17\times0.34$  &   4.0e$-$02  &   5.3e$-$01  &   2.1e$-$04  &   2.9e+07, dis, 3  &   3.2e$-$04  \\
  NGC4579  &  3.13$\pm$0.13e$-07$, 0.26, 3.1e$-$02  &  1.03$\pm$0.02e$-07$, 1.1e$-$01  &  $0.29\times0.59$  &   4.7e$-$02  &   4.7e$-$01  &  \nodata  &   \nodata  &  \nodata  \\
  NGC4593  &  2.65$\pm$0.29e$-07$, 0.04, 1.6e$-$01  &  4.02$\pm$0.38e$-08$, 2.5e$-$01  &  $0.72\times1.44$  &   1.6e$-$01  &   4.8e$-$01  &   1.3e$-$02  &   9.8e+06, rev, 8  &   5.5e$-$02  \\
  NGC4594  &  3.77$\pm$1.09e$-08$, 0.06, 5.3e$-$03  &  6.63$\pm$0.07e$-08$, 9.8e$-$02  &  $0.35\times0.70$  &   3.4e$-$02  &   6.6e$-$01  &   5.3e$-$04  &   5.7e+08, sta, 10  &   4.0e$-$05  \\
  NGC4639  &  1.67$\pm$0.03e$-07$, 0.67, 1.7e$-$02  &  9.46$\pm$0.57e$-09$, 9.9e$-$03  &  $0.29\times0.59$  &   2.1e$-$02  &   1.2e$-$01  &  $<$ 1.9e$-$04  &   7.1e+06, dis, 3  &  $<$ 1.1e$-$03  \\
  NGC4698  &  9.75$\pm$3.34e$-09$, 0.08, 9.7e$-$04  &  5.85$\pm$0.50e$-09$, 6.1e$-$03  &  $0.29\times0.59$  &   5.1e$-$03  &   7.4e$-$02  &  $<$ 3.1e$-$04  &   4.1e+07, dis, 3  &  $<$ 3.3e$-$04  \\
  NGC4725  &  2.83$\pm$1.68e$-08$, 0.06, 1.5e$-$03  &  6.01$\pm$2.35e$-09$, 3.4e$-$03  &  $0.22\times0.43$  &   1.0e$-$02  &   3.4e$-$01  &   1.5e$-$04  &   3.2e+07, dis, 3  &   2.0e$-$04  \\
  NGC4939  &  7.78$\pm$0.94e$-08$, 0.07, 5.9e$-$02  &  6.05$\pm$0.16e$-08$, 4.9e$-$01  &  $0.81\times1.63$  &   5.9e$-$02  &   2.4e+00  &   7.7e$-$02  &  \nodata   &  \nodata   \\
  NGC4941  &  1.31$\pm$0.09e$-07$, 0.08, 1.3e$-$02  &  1.09$\pm$0.01e$-07$, 1.1e$-$01  &  $0.29\times0.59$  &   1.3e$-$02  &   1.2e$-$01  &   4.1e$-$03  &   3.4e+06, dis, 1  &   5.2e$-$02  \\
  NGC4945  &  2.36$\pm$0.02e$-05$,26.11, 1.5e$-$01  &  5.47$\pm$0.05e$-06$, 3.7e$-$01  &  $0.08\times0.15$  &   5.6e$-$01  &   1.1e+00  &  \nodata  &   \nodata  &  \nodata  \\
  NGC5005  &  3.37$\pm$0.05e$-06$, 2.59, 5.4e$-$01  &  3.27$\pm$0.04e$-07$, 5.5e$-$01  &  $0.37\times0.74$  &   6.3e$-$01  &   1.8e+00  &  \nodata  &   \nodata  &  \nodata  \\
  NGC5033  &  1.23$\pm$0.02e$-06$, 1.32, 1.5e$-$01  &  1.27$\pm$0.02e$-07$, 1.6e$-$01  &  $0.33\times0.65$  &   3.3e$-$01  &   1.8e+00  &   1.8e$-$03  &   4.4e+07, dis, 1  &   1.7e$-$03  \\
  NGC5128  &  2.80$\pm$0.28e$-06$, 0.06, 1.8e$-$02  &  2.08$\pm$0.02e$-06$, 1.4e$-$01  &  $0.08\times0.15$  &   3.7e$-$01  &   7.1e$-$01  &   1.4e$-$03  &   7.0e+07, sta, 10  &   8.4e$-$04  \\
  NGC5135  &  4.93$\pm$0.05e$-06$, 1.05, 5.8e+00  &  8.13$\pm$0.08e$-07$, 1.0e+01  &  $1.01\times2.01$  &   6.1e+00  &   3.6e+00  &   1.3e$-$01  &   2.2e+07, dis, 1  &   2.5e$-$01  \\
  NGC5194  &  6.80$\pm$0.15e$-07$, 0.88, 1.7e$-$02  &  1.68$\pm$0.02e$-07$, 4.4e$-$02  &  $0.15\times0.29$  &   2.8e$-$01  &   2.2e+00  &   7.4e$-$04  &   8.9e+06, dis, 1  &   3.5e$-$03  \\
  NGC5273  &  3.20$\pm$0.19e$-07$, 0.63, 5.1e$-$02  &  2.74$\pm$0.30e$-08$, 4.6e$-$02  &  $0.37\times0.74$  &   5.7e$-$02  &   \nodata  &   1.2e$-$03  &   2.1e+06, dis, 1  &   2.5e$-$02  \\
  NGC5395  &  1.22$\pm$0.05e$-07$, 1.30, 9.4e$-$02  &  4.44$\pm$0.50e$-09$, 3.6e$-$02  &  $0.82\times1.63$  &   9.9e$-$02  &   2.4e+00  &  \nodata  &   \nodata  &  \nodata  \\
  NGC5427  &  2.69$\pm$0.04e$-07$, 0.38, 1.5e$-$01  &  4.03$\pm$0.06e$-08$, 2.4e$-$01  &  $0.71\times1.41$  &   2.3e$-$01  &   2.9e+00  &   3.7e$-$03  &  \nodata   &  \nodata   \\
  NGC5506  &  1.82$\pm$0.06e$-06$, 0.10, 5.8e$-$01  &  5.60$\pm$0.06e$-07$, 1.9e+00  &  $0.52\times1.05$  &   5.8e$-$01  &   \nodata  &   1.2e$-$01  &   4.5e+06, dis, 1  &   1.1e+00  \\
  NGC5631  &  2.61$\pm$0.38e$-08$, 0.20, 9.8e$-$03  &  4.41$\pm$0.55e$-09$, 1.7e$-$02  &  $0.57\times1.14$  &   1.3e$-$02  &   3.1e$-$02  &  $<$ 8.6e$-$04  &   6.7e+07, dis, 3  &  $<$ 5.5e$-$04  \\
  NGC5643  &  1.98$\pm$0.09e$-06$, 0.40, 1.4e$-$01  &  2.29$\pm$0.09e$-07$, 1.8e$-$01  &  $0.25\times0.50$  &   2.4e$-$01  &   8.7e$-$01  &   1.4e$-$02  &  \nodata   &  \nodata   \\
  NGC5728  &  2.10$\pm$0.02e$-06$, 1.44, 1.2e+00  &  2.18$\pm$0.02e$-07$, 1.4e+00  &  $0.72\times1.43$  &   1.4e+00  &   1.7e+00  &   1.1e$-$01  &   1.6e+08, dis, 1  &   2.8e$-$02  \\
  NGC5899  &  6.42$\pm$0.32e$-07$, 0.92, 4.1e$-$01  &  8.93$\pm$0.35e$-08$, 6.1e$-$01  &  $0.75\times1.49$  &   4.3e$-$01  &   2.4e+00  &   2.2e$-$02  &  \nodata   &  \nodata   \\
  NGC6221  &  6.50$\pm$0.07e$-06$, 0.83, 8.5e$-$01  &  1.77$\pm$0.02e$-06$, 2.4e+00  &  $0.34\times0.67$  &   1.7e+00  &   3.6e+00  &  \nodata  &  \nodata   &  \nodata   \\
  NGC6300  &  3.07$\pm$0.03e$-06$, 0.34, 2.1e$-$01  &  1.76$\pm$0.02e$-07$, 1.3e$-$01  &  $0.24\times0.49$  &   2.1e$-$01  &   9.3e$-$01  &   3.4e$-$03  &  \nodata   &  \nodata   \\
  NGC6814  &  2.33$\pm$0.21e$-07$, 0.10, 5.4e$-$02  &  6.15$\pm$0.36e$-08$, 1.5e$-$01  &  $0.45\times0.89$  &   9.2e$-$02  &   1.4e+00  &   9.7e$-$03  &   1.8e+07, rev, 11  &   2.2e$-$02  \\
  NGC6951  &  2.03$\pm$0.06e$-06$, 1.59, 4.1e$-$01  &  2.44$\pm$0.04e$-07$, 5.3e$-$01  &  $0.42\times0.84$  &   8.7e$-$01  &   2.0e+00  &   2.6e$-$03  &   2.2e+07, dis, 3  &   5.0e$-$03  \\
  NGC7130  &  2.61$\pm$0.07e$-06$, 0.62, 4.3e+00  &  5.29$\pm$0.08e$-07$, 9.3e+00  &  $1.20\times2.40$  &   4.3e+00  &   6.7e+00  &   3.9e$-$02  &   3.3e+07, dis, 1  &   5.0e$-$02  \\
  NGC7172  &  1.46$\pm$0.06e$-06$, 1.70, 7.3e$-$01  &  2.22$\pm$0.02e$-07$, 1.2e+00  &  $0.66\times1.31$  &   7.9e$-$01  &   6.8e$-$01  &   3.1e$-$02  &   4.7e+07, dis, 1  &   2.8e$-$02  \\
  NGC7213  &  2.04$\pm$0.03e$-07$, 0.06, 4.4e$-$02  &  1.34$\pm$0.03e$-07$, 3.1e$-$01  &  $0.43\times0.87$  &   4.4e$-$02  &   3.9e$-$01  &  \nodata  &  \nodata   &  \nodata   \\
  NGC7314  &  2.55$\pm$0.11e$-07$, 0.12, 3.9e$-$02  &  6.40$\pm$0.14e$-08$, 1.0e$-$01  &  $0.36\times0.73$  &   6.6e$-$02  &   6.2e$-$01  &   1.6e$-$02  &  \nodata   &  \nodata   \\
  NGC7410  &  1.74$\pm$0.13e$-07$, 1.17, 3.8e$-$02  &  2.91$\pm$0.19e$-08$, 6.6e$-$02  &  $0.43\times0.87$  &   4.6e$-$02  &   3.0e$-$01  &  $<$ 4.1e$-$03  &  \nodata   &  \nodata   \\
  NGC7469  &  5.40$\pm$0.05e$-06$, 0.38, 8.5e+00  &  1.17$\pm$0.01e$-06$, 1.9e+01  &  $1.17\times2.34$  &   8.5e+00  &   \nodata  &  \nodata  &   1.2e+07, rev, 5  &  \nodata  \\
  NGC7479  &  8.70$\pm$2.20e$-07$, 0.14, 3.2e$-$01  &  1.16$\pm$0.11e$-07$, 4.5e$-$01  &  $0.57\times1.13$  &   3.2e$-$01  &   1.7e+00  &   2.8e$-$03  &   4.8e+07, dis, 3  &   2.5e$-$03  \\
  NGC7496  &  1.42$\pm$0.03e$-06$, 0.49, 2.7e$-$01  &  2.77$\pm$0.04e$-07$, 5.5e$-$01  &  $0.40\times0.81$  &   2.7e$-$01  &   5.1e$-$01  &  \nodata  &   \nodata  &  \nodata  \\
  NGC7582  &  4.10$\pm$0.04e$-06$, 1.42, 7.0e$-$01  &  3.60$\pm$0.04e$-07$, 6.5e$-$01  &  $0.38\times0.77$  &   2.1e+00  &   1.9e+00  &   5.8e$-$02  &   5.5e+07, gas, 12  &   4.5e$-$02  \\
  NGC7590  &  4.57$\pm$0.81e$-07$, 1.54, 7.8e$-$02  &  2.73$\pm$0.34e$-08$, 4.9e$-$02  &  $0.38\times0.77$  &   1.6e$-$01  &   1.1e+00  &   8.1e$-$04  &   6.2e+06, dis, 1  &   5.6e$-$03  \\
  NGC7743  &  3.42$\pm$0.03e$-07$, 1.23, 7.2e$-$02  &  4.52$\pm$0.28e$-08$, 1.0e$-$01  &  $0.43\times0.85$  &   7.9e$-$02  &   4.9e$-$02  &   9.4e$-$04  &   5.3e+06, dis, 3  &   7.6e$-$03  \\
\enddata

\tablecomments{Col. (1): Galaxy name.  Col. (2): 11.3~$\mu$m aromatic
  feature intensity [W~m$^{-2}$~sr$^{-1}$], equivalent width [$\mu$m],
  and derived star-formation rate [M$_{\odot}$~yr$^{-1}$].  Col. (3):
  \neii\ intensity [W~m$^{-2}$~sr$^{-1}$] and derived star-formation
  rate [M$_{\odot}$~yr$^{-1}$].  Col. (4): Physical size of the
  $3.6\arcsec\times7.2\arcsec$ aperture used for the two previous
  columns [kpc$\times$kpc].  Col. (5): Star-formation rate inside
  $r=1$~kpc [M$_{\odot}$~yr$^{-1}$].  Col. (6): Star-formation rate
  outside $r=1$~kpc [M$_{\odot}$~yr$^{-1}$].  Col. (7): Black hole
  accretion rate dervied from the \oiv\ luminosity
  [M$_{\odot}$~yr$^{-1}$].  Col. (8): Black hole mass [M$_{\odot}$],
  method used to determine black hole mass (mas: maser dynamics, dis:
  bulge velocity dispersion, rev: reverberation mapping, gas: gas
  dynamics, ste: stellar dynamics), and reference for black hole mass
  (see below).  Col. (9): Eddington ratio.  References: (1)
  \citet{bia07}.  (2) \citet{pet04}.  (3) \citet{ho09}.  (4)
  \citet{dev03}.  (5) \citet{den09}.  (6) \citet{ben06}.  (7)
  \citet{her05}.  (8) \citet{den06}.  (9) \citet{kor88}.  (10)
  \citet{cap09}.  (11) \citet{ben09}.  (12) \citet{wol06}.}

\end{deluxetable}

\end{landscape}

\begin{deluxetable}{lllllllll}
\tabletypesize{\scriptsize}
\tablecaption{Correlation Analysis: SFR v. $\dot{M}_{BH}$ \label{t2}}
\tablewidth{0pt}
\tablehead{
\colhead{aperture} & \colhead{11.3~$\mu$m} & {24~$\mu$m} &  \colhead{$\alpha$} & \colhead{$\beta$} & \colhead{$\sigma$} & \colhead{luminosity} & \colhead{flux} & \colhead{flux} \\ 
\colhead{radius}         & \colhead{PAH} &  \colhead{continuum}  &  \colhead{}         & \colhead{}         & \colhead{}         & \colhead{correlation} & \colhead{correlation} & \colhead{probability} } 
\startdata 
 $<$300~pc$>$    & Y & N  &$0.88^{+0.36}_{-0.31}$ & $0.80^{+0.14}_{-0.12}$ & $0.37^{+0.18}_{-0.20}$ & $0.95^{+0.04}_{-0.07}$ & $0.69^{+0.11}_{-0.13}$ & $<1\times10^{-6}$\\
 $1$~kpc     & Y & Y & $0.67^{+0.39}_{-0.29}$ & $0.61^{+0.15}_{-0.11}$ & $0.41^{+0.23}_{-0.24}$ & $0.93^{+0.06}_{-0.10}$ & $0.49^{+0.16}_{-0.18}$ & $2.9\times10^{-5}$\\
 $>1$~kpc    & N & Y  &$1.30^{+0.76}_{-0.45}$ & $0.57^{+0.28}_{-0.17}$ & $0.73^{+0.13}_{-0.20}$ & $0.78^{+0.13}_{-0.20}$ & $0.33^{+0.21}_{-0.22}$ & 0.02 \\
 $r_{galaxy}$ & Y & Y  &$1.59^{+1.15}_{-0.57}$ & $0.71^{+0.45}_{-0.22}$ & $0.86^{+0.22}_{-0.20}$ & $0.65^{+0.17}_{-0.22}$ & $0.14^{+0.20}_{-0.21}$ & 0.39 \\ 
\enddata
\tablecomments{$\log(SFR)=\alpha+\log(\dot{M}_{BH})\times\beta\pm\sigma$}
\tablecomments{The quoted values in columns 4--8 correspond to the median of the posterior distribution, and the range that encompasses 90\% of that distribution \citep[see][]{kel07}.  The values in the last column correspond to the probability that no correlation exists in the flux-flux version of the relationship, based on generalized Spearman's $\rho$ \citep{iso86,lav92}.}
\end{deluxetable}


\begin{thebibliography}{}
\bibitem[Abramowicz et al.(1988)]{abr88} Abramowicz, M.~A., Czerny,
  B., Lasota, J.~P., \& Szuszkiewicz, E.\ 1988, \apj, 332, 646
\bibitem[Alexander et al.(2003)]{ale03} Alexander, D.~M., et
  al.\ 2003, \aj, 126, 539
\bibitem[Antonucci(1993)]{ant93} Antonucci, R.\ 1993, \araa, 31, 473
\bibitem[Assef et al.(2010)]{ass10} Assef, R.~J., et al.\ 2010, \apj,
  713, 970
\bibitem[Atlee et al.(2011)]{atl11} Atlee, D.~W., Martini, P., Assef,
  R.~J., Kelson, D.~D., \& Mulchaey, J.~S.\ 2011, \apj, 729, 22
\bibitem[Ballantyne et al.(2006)]{bal06} Ballantyne, D.~R., Everett,
  J.~E., \& Murray, N.\ 2006, \apj, 639, 740
\bibitem[Ballantyne(2008)]{ball08} Ballantyne, D.~R.\ 2008, \apj, 685,
  787
\bibitem[Balmaverde et al.(2008)]{bal08} Balmaverde, B., Baldi, R.~D.,
  \& Capetti, A.\ 2008, \aap, 486, 119
\bibitem[Barnes \& Hernquist(1991)]{bar91} Barnes, J.~E., \&
  Hernquist, L.~E.\ 1991, \apjl, 370, L65
\bibitem[Bentz et al.(2006)]{ben06} Bentz, M.~C., et al.\ 2006, \apj,
  651, 775
\bibitem[Bentz et al.(2009)]{ben09} Bentz, M.~C., et al.\ 2009, \apj,
  705, 199
\bibitem[Bian \& Gu(2007)]{bia07} Bian, W., \& Gu, Q.\ 2007, \apj,
  657, 159
\bibitem[Bonfield et al.(2011)]{bon11} Bonfield, D.~G., et al.\ 2011,
  \mnras, arXiv:1103.3905
\bibitem[Brand et al.(2006)]{bra06} Brand, K., et al.\ 2006, \apj,
  644, 143
\bibitem[Brinchmann et al.(2004)]{bri04} Brinchmann, J., Charlot, S.,
  White, S.~D.~M., Tremonti, C., Kauffmann, G., Heckman, T., \&
  Brinkmann, J.\ 2004, \mnras, 351, 1151
\bibitem[Buchanan et al.(2006)]{buc06} Buchanan, C.~L., Gallimore,
  J.~F., O'Dea, C.~P., Baum, S.~A., Axon, D.~J., Robinson, A.,
  Elitzur, M., \& Elvis, M.\ 2006, \aj, 132, 401
\bibitem[Calzetti et al.(2007)]{cal07} Calzetti, D., et al.\ 2007,
  \apj, 666, 870
\bibitem[Cappellari et al.(2009)]{cap09} Cappellari, M., Neumayer, N.,
  Reunanen, J., van der Werf, P.~P., de Zeeuw, P.~T., \& Rix,
  H.-W.\ 2009, \mnras, 394, 660
\bibitem[Cid Fernandes et al.(2001)]{cid01} Cid Fernandes, R.,
  Heckman, T., Schmitt, H., Gonz{\'a}lez Delgado, R.~M., \&
  Storchi-Bergmann, T.\ 2001, \apj, 558, 81
\bibitem[Ciotti \& Ostriker(2007)]{cio07} Ciotti, L., \& Ostriker,
  J.~P.\ 2007, \apj, 665, 1038
\bibitem[Cisternas et al.(2011)]{cis11} Cisternas, M., et al.\ 2011,
  \apj, 726, 57
\bibitem[Cusumano et al.(2010)]{cus10} Cusumano, G., et al.\ 2010,
  \aap, 510, A48
\bibitem[Davies et al.(2007)]{dav07} Davies, R.~I., S{\'a}nchez,
  F.~M., Genzel, R., Tacconi, L.~J., Hicks, E.~K.~S., Friedrich, S.,
  \& Sternberg, A.\ 2007, \apj, 671, 1388
\bibitem[Denney et al.(2006)]{den06} Denney, K.~D., et al.\ 2006,
  \apj, 653, 152
\bibitem[Denney et al.(2009)]{den09} Denney, K.~D., et al.\ 2009,
  \apj, 702, 1353
\bibitem[Deo et al.(2007)]{deo07} Deo, R.~P., Crenshaw, D.~M.,
  Kraemer, S.~B., Dietrich, M., Elitzur, M., Teplitz, H., \& Turner,
  T.~J.\ 2007, \apj, 671, 124
\bibitem[Devereux et al.(2003)]{dev03} Devereux, N., Ford, H.,
  Tsvetanov, Z., \& Jacoby, G.\ 2003, \aj, 125, 1226
\bibitem[Diamond-Stanic et al.(2009)]{dia09} Diamond-Stanic, A.~M.,
Rieke, G.~H., \& Rigby, J.~R.\ 2009, \apj, 698, 623
\bibitem[Diamond-Stanic \& Rieke(2010)]{dia10} Diamond-Stanic, A.~M.,
  \& Rieke, G.~H.\ 2010, \apj, 724, 140
\bibitem[D{\'{\i}}az-Santos et al.(2010)]{diaz10} D{\'{\i}}az-Santos,
  T., Alonso-Herrero, A., Colina, L., Packham, C., Levenson, N.~A.,
  Pereira-Santaella, M., Roche, P.~F., \& Telesco, C.~M.\ 2010, \apj,
  711, 328
\bibitem[Di Matteo et al.(2005)]{dim05} Di Matteo, T., Springel, V.,
  \& Hernquist, L.\ 2005, \nat, 433, 604
\bibitem[Dudik et al.(2005)]{dud05} Dudik, R.~P., Satyapal, S.,
  Gliozzi, M., \& Sambruna, R.~M.\ 2005, \apj, 620, 113
\bibitem[Escala(2007)]{esc07} Escala, A.\ 2007, \apj, 671, 1264
\bibitem[Ferrarese \& Merritt(2000)]{fer00} Ferrarese, L., \& Merritt,
  D.\ 2000, \apjl, 539, L9
\bibitem[Fu et al.(2010)]{fu10} Fu, H., et al.\ 2010, \apj, 722, 653
\bibitem[Gabor et al.(2009)]{gab09} Gabor, J.~M., et al.\ 2009, \apj,
  691, 705
\bibitem[Gebhardt et al.(2000)]{geb00} Gebhardt, K., et al.\ 2000,
  \apjl, 539, L13
\bibitem[Genzel et al.(1998)]{gen98} Genzel, R., et al.\ 1998, \apj,
  498, 579
\bibitem[Graham(2008)]{gra08a} Graham, A.~W.\ 2008, \apj, 680, 143
\bibitem[Graham \& Worley(2008)]{gra08b} Graham, A.~W., \& Worley,
  C.~C.\ 2008, \mnras, 388, 1708
\bibitem[Greene et al.(2010)]{gre10} Greene, J.~E., et al.\ 2010,
  \apj, 721, 26
\bibitem[Grogin et al.(2005)]{gro05} Grogin, N.~A., et al.\ 2005,
  \apjl, 627, L97
\bibitem[Groves et al.(2006)]{gro06} Groves, B., Dopita, M., \&
  Sutherland, R.\ 2006, \aap, 458, 405
\bibitem[H{\"a}ring \& Rix(2004)]{har04} H{\"a}ring, N., \& Rix,
  H.-W.\ 2004, \apjl, 604, L89
\bibitem[Hatziminaoglou et al.(2010)]{hat10} Hatziminaoglou, E., et
  al.\ 2010, \aap, 518, L33
\bibitem[Hao et al.(2005)]{hao05} Hao, C.~N., Xia, X.~Y., Mao, S., Wu,
  H., \& Deng, Z.~G.\ 2005, \apj, 625, 78
\bibitem[Hao et al.(2009)]{hao09} Hao, L., Jogee, S., Barazza, F.~D.,
  Marinova, I., \& Shen, J.\ 2009, Galaxy Evolution: Emerging Insights
  and Future Challenges, 419, 402
\bibitem[Heckman et al.(2004)]{hec04} Heckman, T.~M., Kauffmann, G.,
  Brinchmann, J., Charlot, S., Tremonti, C., \& White, S.~D.~M.\ 2004,
  \apj, 613, 109
\bibitem[Herrnstein et al.(2005)]{her05} Herrnstein, J.~R., Moran,
  J.~M., Greenhill, L.~J., \& Trotter, A.~S.\ 2005, \apj, 629, 719
\bibitem[Ho et al.(1997a)]{ho97} Ho, L.~C., Filippenko, A.~V., \&
  Sargent, W.~L.~W.\ 1997, \apjs, 112, 315
\bibitem[Ho et al.(1997b)]{ho97b} Ho, L.~C., Filippenko, A.~V., \&
  Sargent, W.~L.~W.\ 1997, \apj, 487, 591
\bibitem[Ho \& Keto(2007)]{ho07} Ho, L.~C., \& Keto, E.\ 2007, \apj,
  658, 314
\bibitem[Ho et al.(2009)]{ho09} Ho, L.~C., Greene, J.~E., Filippenko,
  A.~V., \& Sargent, W.~L.~W.\ 2009, \apjs, 183, 1
\bibitem[Hobbs et al.(2011)]{hob11} Hobbs, A., Nayakshin, S., Power,
  C., \& King, A.\ 2011, \mnras, 413, 2633
\bibitem[Hopkins et al.(2005)]{hop05} Hopkins, P.~F., Hernquist, L.,
  Cox, T.~J., Di Matteo, T., Martini, P., Robertson, B., \& Springel,
  V.\ 2005, \apj, 630, 705
\bibitem[Hopkins \& Hernquist(2006)]{hop06} Hopkins, P.~F., \& Hernquist,
  L.\ 2006b, \apjs, 166, 1
\bibitem[Hopkins \& Quataert(2010)]{hop10} Hopkins, P.~F., \&
  Quataert, E.\ 2010, \mnras, 1085
\bibitem[Houck et al.(2004)]{hou04} Houck, J.~R., et al.\ 2004, \apjs,
154, 18
\bibitem[Hu(2008)]{hu08} Hu, J.\ 2008, \mnras, 386, 2242
\bibitem[Hunt \& Malkan(1999)]{hun99} Hunt, L.~K., \& Malkan,
  M.~A.\ 1999, \apj, 516, 660
\bibitem[Huchra \& Burg(1992)]{huc92} Huchra, J., \& Burg, R.\ 1992,
  \apj, 393, 90
\bibitem[Imanishi \& Wada(2004)]{ima04} Imanishi, M., \& Wada,
  K.\ 2004, \apj, 617, 214
\bibitem[Isobe et al.(1986)]{iso86} Isobe, T., Feigelson, E.~D., \&
  Nelson, P.~I.\ 1986, \apj, 306, 490
\bibitem[Jogee(2006)]{jog06} Jogee, S.\ 2006, Physics of Active
  Galactic Nuclei at all Scales, 693, 143
\bibitem[Kauffmann \& Haehnelt(2000)]{kau00} Kauffmann, G., \&
  Haehnelt, M.\ 2000, \mnras, 311, 576
\bibitem[Kauffmann et al.(2003)]{kau03} Kauffmann, G., et al.\ 2003,
  \mnras, 346, 1055
\bibitem[Kawakatu \& Wada(2008)]{kaw08} Kawakatu, N., \& Wada,
  K.\ 2008, \apj, 681, 73
\bibitem[Kelly(2007)]{kel07} Kelly, B.~C.\ 2007, \apj, 665, 1489
\bibitem[Kennicutt(1998a)]{ken98} Kennicutt, R.~C., Jr.\ 1998, \araa,
  36, 189
\bibitem[Kennicutt(1998b)]{ken98b} Kennicutt, R.~C., Jr.\ 1998, \apj,
  498, 541
\bibitem[Kormendy(1988)]{kor88} Kormendy, J.\ 1988, \apj, 335, 40
\bibitem[Kormendy \& Richstone(1995)]{kor95} Kormendy, J., \&
  Richstone, D.\ 1995, \araa, 33, 581
\bibitem[Kormendy et al.(2011)]{kor11} Kormendy, J., Bender, R., \&
  Cornell, M.~E.\ 2011, \nat, 469, 374
\bibitem[Laurikainen et al.(2004)]{lau04} Laurikainen, E., Salo, H.,
  \& Buta, R.\ 2004, \apj, 607, 103
\bibitem[Lavalley et al.(1992)]{lav92} Lavalley, M., Isobe, T., \&
  Feigelson, E.\ 1992, Astronomical Data Analysis Software and Systems
  I, 25, 245
\bibitem[Lehmer et al.(2005)]{leh05} Lehmer, B.~D., et al.\ 2005,
  \apjs, 161, 21
\bibitem[Lutz et al.(2008)]{lut08} Lutz, D., et al.\ 2008, \apj, 684,
  853
\bibitem[Lutz et al.(2010)]{lut10} Lutz, D., et al.\ 2010, \apj, 712,
  1287
\bibitem[Maiolino et al.(1995)]{mai95b} Maiolino, R.,
  Ruiz, M., Rieke, G.~H., \& Keller, L.~D.\ 1995, \apj, 446, 561
\bibitem[Maiolino \& Rieke(1995)]{mai95} Maiolino, R., \& Rieke,
G.~H.\ 1995, \apj, 454, 95
\bibitem[Magorrian et al.(1998)]{mag98} Magorrian, J., et al.\ 1998,
  \aj, 115, 2285
\bibitem[Marconi \& Hunt(2003)]{mar03} Marconi, A., \&
  Hunt, L.~K.\ 2003, \apjl, 589, L21
\bibitem[Mel{\'e}ndez et al.(2008a)]{mel08a} Mel{\'e}ndez, M., et
  al.\ 2008a, \apj, 682, 94
\bibitem[Mel{\'e}ndez et al.(2008b)]{mel08b} Mel{\'e}ndez, M., Kraemer,
  S.~B., Schmitt, H.~R., Crenshaw, D.~M., Deo, R.~P., Mushotzky,
  R.~F., \& Bruhweiler, F.~C.\ 2008, \apj, 689, 95
\bibitem[Miller et al.(2010)]{mil10} Miller, J.~M., Nowak, M.,
  Markoff, S., Rupen, M.~P., \& Maitra, D.\ 2010, \apj, 720, 1033
\bibitem[Mouri \& Taniguchi(2002)]{mou02} Mouri, H., \& Taniguchi,
  Y.\ 2002, \apj, 565, 786
\bibitem[Mulchaey \& Regan(1997)]{mul97} Mulchaey, J.~S., \& Regan,
  M.~W.\ 1997, \apjl, 482, L135
\bibitem[Mullaney et al.(2011)]{mul11} Mullaney, J.~R., Pannella, M.,
  Daddi, E., et al.\ 2011, \mnras, 1756, arXiv:1106.4284
\bibitem[Narayan \& Yi(1995)]{nar95} Narayan, R., \& Yi, I.\ 1995,
  \apj, 452, 710
\bibitem[Netzer et al.(2007)]{net07} Netzer, H., et al.\ 2007, \apj,
  666, 806
\bibitem[Netzer(2009)]{net09} Netzer, H.\ 2009, \mnras, 399, 1907
\bibitem[Norman \& Scoville(1988)]{nor88} Norman, C., \& Scoville,
  N.\ 1988, \apj, 332, 124
\bibitem[O'Dowd et al.(2009)]{odo09} O'Dowd, M.~J., et al.\ 2009,
  \apj, 705, 885
\bibitem[Pierce et al.(2007)]{pie07} Pierce, C.~M., et al.\ 2007,
  \apjl, 660, L19
\bibitem[Pilbratt et al.(2010)]{pil10} Pilbratt, G.~L., et al.\ 2010,
  \aap, 518, L1
\bibitem[Peeters et al.(2004)]{pee04} Peeters, E., Spoon, H.~W.~W., \&
  Tielens, A.~G.~G.~M.\ 2004, \apj, 613, 986
\bibitem[Peterson et al.(2004)]{pet04} Peterson, B.~M., et al.\ 2004,
  \apj, 613, 682
\bibitem[Pereira-Santaella et al.(2010)]{per10} Pereira-Santaella, M.,
  Diamond-Stanic, A.~M., Alonso-Herrero, A., \& Rieke, G.~H.\ 2010,
  \apj, 725, 2270
\bibitem[Quataert et al.(1999)]{qua99} Quataert, E., Di Matteo, T.,
  Narayan, R., \& Ho, L.~C.\ 1999, \apjl, 525, L89
\bibitem[Rieke et al.(2004)]{rie04} Rieke, G.~H., et al.\ 2004, \apjs,
  154, 25
\bibitem[Rieke et al.(2009)]{rie09} Rieke, G.~H., Alonso-Herrero, A.,
Weiner, B.~J., P{\'e}rez-Gonz{\'a}lez, P.~G., Blaylock, M., Donley,
J.~L., \& Marcillac, D.\ 2009, \apj, 692, 556
\bibitem[Riffel et al.(2009)]{rif09} Riffel, R.,
  Pastoriza, M.~G., Rodr{\'{\i}}guez-Ardila, A., \& Bonatto, C.\ 2009,
  \mnras, 400, 273
\bibitem[Rigby et al.(2009)]{rig09} Rigby, J.~R., Diamond-Stanic,
  A.~M., \& Aniano, G.\ 2009, \apj, 700, 1878
\bibitem[Rush et al.(1993)]{rus93} Rush, B., Malkan, M.~A., \&
  Spinoglio, L.\ 1993, \apjs, 89, 1
\bibitem[Sandage \& Tammann(1987)]{san87} Sandage, A., \& Tammann,
G.~A.\ 1987, A Revised Shapley--Ames Catalog of Bright Galaxies (2nd
ed. Washington, DC: Carnegie Institution of Washington)
\bibitem[Sanders et al.(1988)]{san88} Sanders, D.~B., Soifer, B.~T.,
  Elias, J.~H., Madore, B.~F., Matthews, K., Neugebauer, G., \&
  Scoville, N.~Z.\ 1988, \apj, 325, 74
\bibitem[Satyapal et al.(2005)]{sat05} Satyapal, S., Dudik, R.~P.,
  O'Halloran, B., \& Gliozzi, M.\ 2005, \apj, 633, 86
\bibitem[Serjeant \& Hatziminaoglou(2009)]{ser09} Serjeant, S., \&
  Hatziminaoglou, E.\ 2009, \mnras, 397, 265
\bibitem[Schmidt(1959)]{sch59} Schmidt, M.\ 1959, \apj, 129, 243
\bibitem[Schweitzer et al.(2006)]{sch06} Schweitzer, M., et al.\ 2006,
  \apj, 649, 79
\bibitem[Scoville et al.(2007)]{sco07} Scoville, N., et al.\ 2007,
  \apjs, 172, 1
\bibitem[Shao et al.(2010)]{sha10} Shao, L., et al.\ 2010, \aap, 518,
  L26
\bibitem[Shen et al.(2003)]{she03} Shen, S., Mo, H.~J., White,
  S.~D.~M., Blanton, M.~R., Kauffmann, G., Voges, W., Brinkmann, J.,
  \& Csabai, I.\ 2003, \mnras, 343, 978
\bibitem[Shi et al.(2007)]{shi07} Shi, Y., et al.\ 2007, \apj, 669,
841
\bibitem[Shi et al.(2009)]{shi09} Shi, Y., Rieke, G.~H., Ogle, P.,
  Jiang, L., \& Diamond-Stanic, A.~M.\ 2009, \apj, 703, 1107
\bibitem[Shlosman et al.(1990)]{shl90} Shlosman, I., Begelman, M.~C.,
  \& Frank, J.\ 1990, \nat, 345, 679
\bibitem[Silverman et al.(2009)]{sil09} Silverman, J.~D., et
  al.\ 2009, \apj, 696, 396
\bibitem[Silk \& Rees(1998)]{sil98} Silk, J., \& Rees, M.~J.\ 1998,
  \aap, 331, L1
\bibitem[Smith et al.(2007a)]{smi07a} Smith, J.~D.~T., et al.\ 2007,
\apj, 656, 770
\bibitem[Smith et al.(2007b)]{smi07b} Smith, J.~D.~T., et al.\ 2007,
\pasp, 119, 1133
\bibitem[Smith et al.(2010)]{smi10} Smith, H.~A., Li, A., Li, M.~P.,
  et al.\ 2010, \apj, 716, 490
\bibitem[Somerville et al.(2008)]{som08} Somerville, R.~S., Hopkins,
  P.~F., Cox, T.~J., Robertson, B.~E., \& Hernquist, L.\ 2008, \mnras,
  391, 481
\bibitem[Storchi-Bergmann et al.(2001)]{sto01} Storchi-Bergmann, T.,
  Gonz{\'a}lez Delgado, R.~M., Schmitt, H.~R., Cid Fernandes, R., \&
  Heckman, T.\ 2001, \apj, 559, 147
\bibitem[Sturm et al.(2002)]{stu02} Sturm, E., Lutz, D., Verma, A., et
  al.\ 2002, \aap, 393, 821
\bibitem[Thompson et al.(2005)]{tho05} Thompson, T.~A., Quataert, E.,
  \& Murray, N.\ 2005, \apj, 630, 167
\bibitem[Tielens(2005)]{tie05} Tielens, A.~G.~G.~M.\ 2005, The Physics
  and Chemistry of the Interstellar Medium (Cambridge: Cambridge
  Univ. Press)
\bibitem[Tommasin et al.(2010)]{tom10} Tommasin, S., Spinoglio, L.,
  Malkan, M.~A., \& Fazio, G.\ 2010, \apj, 709, 1257
\bibitem[Tozzi et al.(2006)]{toz06} Tozzi, P., et al.\ 2006, \aap,
  451, 457
\bibitem[Tremaine et al.(2002)]{tre02} Tremaine, S., et al.\ 2002,
  \apj, 574, 740
\bibitem[Treyer et al.(2010)]{tre10} Treyer, M., et al.\ 2010, \apj,
  719, 1191
\bibitem[Urry \& Padovani(1995)]{urr95} Urry, C.~M., \& Padovani,
  P.\ 1995, \pasp, 107, 803
\bibitem[Voit(1992)]{voi92} Voit, G.~M.\ 1992, \mnras, 258, 841
\bibitem[von Linden et al.(1993)]{von93} von Linden, S., Biermann,
  P.~L., Duschl, W.~J., Lesch, H., \& Schmutzler, T.\ 1993, \aap, 280,
  468
\bibitem[Wada \& Norman(2002)]{wad02} Wada, K., \& Norman,
  C.~A.\ 2002, \apjl, 566, L21
\bibitem[Wada(2004)]{wad04} Wada, K. 2004, Carnegie Observatories
  Astrophys. Ser. 1, Coevolution of Black Holes and Galaxies,
  ed. L. C. Ho (Cambridge: Cambridge Univ. Press), 186
\bibitem[Weaver et al.(2010)]{wea10} Weaver, K.~A., Mel{\'e}ndez, M.,
  Mushotzky, R.~F., et al.\ 2010, \apj, 716, 1151
\bibitem[Werner et al.(2004)]{wer04} Werner, M.~W., et al.\ 2004,
  \apjs, 154, 1
\bibitem[Wild et al.(2007)]{wil07} Wild, V., Kauffmann, G., Heckman,
  T., Charlot, S., Lemson, G., Brinchmann, J., Reichard, T., \&
  Pasquali, A.\ 2007, \mnras, 381, 543
\bibitem[Wild et al.(2010)]{wil10} Wild, V., Heckman, T., \& Charlot,
  S.\ 2010, \mnras, 405, 933
\bibitem[Wold et al.(2006)]{wol06} Wold, M., Lacy, M., K{\"a}ufl,
  H.~U., \& Siebenmorgen, R.\ 2006, \aap, 460, 449
\bibitem[Wu et al.(2006)]{wu06} Wu, Y., Charmandaris, V., Hao, L.,
  Brandl, B.~R., Bernard-Salas, J., Spoon, H.~W.~W., \& Houck,
  J.~R.\ 2006, \apj, 639, 157
\bibitem[Wyithe \& Loeb(2003)]{wyi03} Wyithe, J.~S.~B., \& Loeb,
  A.\ 2003, \apj, 595, 614
\end{thebibliography}
\end{document}